%% file: ms.tex
\documentclass[conference,10pt,twoside,twocolumn]{IEEEtran}

\usepackage{amsmath,amsfonts}
\usepackage{array}
\usepackage[caption=false,font=normalsize,labelfont=sf,textfont=sf]{subfig}
\usepackage{textcomp}
\usepackage{stfloats}
\usepackage{url}
\usepackage{verbatim}
\usepackage{graphicx}
\usepackage{cite}
\usepackage{nicefrac}
\usepackage[dvipsnames]{xcolor}
\usepackage{balance}

\usepackage{twemojis}

\usepackage[hidelinks]{hyperref}
\hypersetup{
    colorlinks=false,
    linkcolor=black
    }

\usepackage{tcolorbox}
\usepackage{bm}			
\usepackage[per-mode=symbol]{siunitx}    		
\usepackage{cite} 
\DeclareSIUnit{\dBm}{dBm}	
\DeclareSIUnit{\eq}{eq}	    
\DeclareSIUnit{\sqrtW}{\ensuremath{\sqrt{\text{W}}}}
\usepackage{layouts}
\usepackage{overpic}		
\usepackage{array, booktabs, xltabular, multirow} 
\usepackage{amssymb}    
\usepackage[capitalise]{cleveref}

\definecolor{RDlightgreen}{RGB}{141 192 69}
\definecolor{RDgreen}{RGB}{93 109 68}
\definecolor{mygray}{gray}{0.6}
                
\usepackage{enumerate}  
\usepackage[xindy]{glossaries}
\makenoidxglossaries%
\glsdisablehyper
\loadglsentries{abbr}%

\usepackage{pgfplots}

\pgfplotsset{compat=newest}
\usetikzlibrary{plotmarks}
\usetikzlibrary{arrows.meta}
\usetikzlibrary{calc}
\usepackage{tikzscale}		
\usepackage{orcidlink}
\usetikzlibrary{patterns}
\usetikzlibrary{positioning,spy}
\usetikzlibrary{scopes}
\usetikzlibrary{backgrounds}

\usepackage{algorithm2e}		
\RestyleAlgo{ruled}			
\SetKw{KwBy}{by}			%
\SetKwComment{Comment}{\% }{}	

\usepackage[hang,flushmargin]{footmisc}
\makeatletter
\newcommand{\algorithmfootnote}[2][\footnotesize]{%
  \let\old@algocf@finish\@algocf@finish
  \def\@algocf@finish{\old@algocf@finish
    \leavevmode\rlap{\begin{minipage}{\linewidth}
    #1#2
    \end{minipage}}%
    \vspace{-0.3cm} 
  }%
}

\definecolor{IEEEblue}{RGB}{0 98 155}
\definecolor{IEEElightblue}{RGB}{0 181 226}
\definecolor{IEEEturquoise}{RGB}{0 156 166}
\definecolor{IEEEred}{RGB}{186 12 47}
\definecolor{IEEEgreen}{RGB}{0 132 61}
\definecolor{IEEElightgreen}{RGB}{120 190 32}
\definecolor{IEEEorange}{RGB}{225 163 0}
\definecolor{IEEEyellow}{RGB}{255 209 0}

\newcommand{\herm}{\mathsf{H}}
\newcommand{\trp}{\mathsf{T}}

\DeclareMathOperator{\diag}{diag}
\DeclareMathOperator*{\argmax}{arg\,max}

\DeclareMathOperator{\vectorize}{vec}
\newcommand{\complexset}[2]{ \mathbb{C}^{#1 \times #2}  }

\newlength\figureheight
\newlength\figurewidth 

\newcounter{romanNumerals}
\input{macros/vmr-symbols-vecbold}

\IEEEoverridecommandlockouts

\newcommand{\customlineref}[1]{
    \tikz[baseline={([yshift=-.5ex]current bounding box.center)}, inner sep=0, outer sep=0,xshift =0pt]{
        \hspace*{-0.2em} 
        \draw[#1] (0,0) -- (0.6cm,0);
    }\hspace*{-0.4em}
}

\begin{document}
\input{math-notations}
\input{macros/colorsGian}

\title{Accurate Direct Positioning in Distributed MIMO Using Delay-Doppler Channel Measurements}

\author{
\IEEEauthorblockN{
    Benjamin Deutschmann\textsuperscript{\twemoji{austria}},
    Christian Nelson\textsuperscript{\twemoji{sweden}},
    Mikael Henriksson\textsuperscript{\twemoji{lion}},
    Gian Marti\textsuperscript{\twemoji{switzerland}}, \\
    Alva Kosasih\textsuperscript{\twemoji{crown}},
    Nuutti Tervo\textsuperscript{\twemoji{finland}},
    Erik Leitinger\textsuperscript{\twemoji{austria}},
    and Fredrik Tufvesson\textsuperscript{\twemoji{sweden}}
    \vspace{1mm}
}
\IEEEauthorblockA{
    \small
    \textsuperscript{\twemoji{austria}}TU Graz, 
    \textsuperscript{\twemoji{sweden}}Lund University,
    \textsuperscript{\twemoji{lion}}Link\"oping University,
    \textsuperscript{\twemoji{switzerland}}ETH Zurich,
    \textsuperscript{\twemoji{crown}}KTH Royal Institute of Technology,
    \textsuperscript{\twemoji{finland}}University of Oulu
    \vspace{1mm}\\
    email: benjamin.deutschmann@tugraz.at, christian.nelson@eit.lth.se, mikael.henriksson@liu.se, marti@iis.ee.ethz.ch, \\ kosasih@kth.se, nuutti.tervo@oulu.fi, erik.leitinger@tugraz.at, fredrik.tufvesson@eit.lth.se
}
\thanks{
    This work is supported by the Excellence Center at Link\"oping -- Lund in Information Technology (ELLIIT) and has 
    been done in part during the 2023 ELLIIT focus period ``6G -- forming a better future.'' The project has received funding from 
the Christian Doppler Research Association and 
the European Union's Horizon 2020 research and innovation program under grant agreement No 101013425 (project ``REINDEER'').
Emojis by Twitter, Inc.\! and others are licensed under \mbox{CC-BY 4.0}.}
}

\maketitle
\begin{abstract}
Distributed multiple-input multiple-output (D-MIMO) is a promising technology for simultaneous communication and positioning. However, phase synchronization between multiple access points in D-MIMO is challenging and methods that function without the need for phase synchronization are highly desired. 
Therefore, we present a method for D-MIMO that performs direct positioning of a moving device based on the delay-Doppler characteristics of the channel state information~(CSI). Our method relies on particle-filter-based Bayesian inference with a state-space model. We use recent measurements from a sub-6~GHz D-MIMO OFDM system in an industrial environment to demonstrate near-centimeter accuracy under partial line-of-sight (LoS) conditions and decimeter accuracy under fully obstructed LoS. 
\end{abstract}

\glsresetall

\section{Introduction}
Next-generation wireless systems are envisioned to integrate communications with sensing capabilities such as
positioning \cite{Liu2022IntegratedSensing, Behravan2023Positioning}
and they are expected to employ geographically separated antennas or \glspl{ap}~\cite{Wang2013DMIMO, Nayebi2015CellFree}. 
Compared to traditional co-located massive \gls{mimo} technology, 
distributed \glspl{ap} provide significantly increased spatial diversity, which improves 
coverage throughout the service area and enables more accurate positioning \cite{Fascista2023, Liu2022IntegratedSensing}.
However, in wireless systems with distributed APs, calibration, and synchronization cannot be taken for granted~\cite{Larsson2024Synchrony}.
Next-generation integrated communications and positioning therefore require the development of robust algorithms that accommodate the limitations of real-world wireless systems. 

\subsection{Contributions}
\begin{figure}[tp]
\centering
\setlength{\figurewidth}{0.99\linewidth}
\input{figures/scenarios_v2}
\vspace{-0.2cm}\caption{Floorplan of the industrial hall used for positioning. 
The floorplan depicts major obstacles (in gray), the twelve distributed receive antennas (in red),  
and the measured tracks used for positioning (in different shades of~blue). The light gray area indicates a region dominated by obstructed conditions mainly due to a lower ceiling height.}
\label{fig:scenarios}
\end{figure}
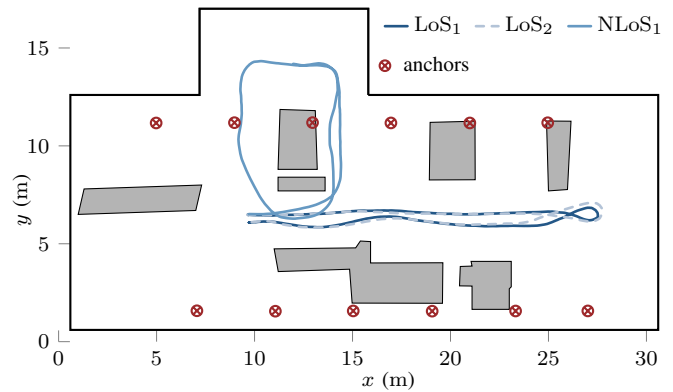

We present a method for direct positioning in \gls{dmimo} communication systems based on delay-Doppler characteristics of \gls{ofdm} 
channel measurements. Our method estimates the position of a moving device, called \emph{agent,} through particle-filter-based 
Bayesian inference using a state-space model where the observations consist of estimates of the \gls{csi} between the agent and a set of distributed stationary APs, called \emph{anchors.} No phase synchronization between the anchors is assumed. We evaluate the method on recent measurements of a sub-6 GHz OFDM system~\cite{Nelson2024} between the moving agent and 12 distributed single-antenna anchors located in a $\SI{30}{\metre} \times \SI{12}{\metre}$ industrial environment, see Fig.\,\ref{fig:scenarios}.
Despite limited bandwidth and no phase synchronization, the results show that our method achieves near-centimeter-level accuracy under partial \gls{los} conditions, and decimeter accuracy under fully \gls{olos} conditions. 

\newcommand{\trajectoryLW}{1.0pt}     

\subsection{Related Work}
The potential of (D-)MIMO CSI for integrated positioning and communications has long been recognized. 
Reference \cite{Chen2022a} performs positioning based on Doppler shifts in co-located MIMO. 
References \cite{yang2023multiple} and \cite{blanco2022augmenting} perform positioning in \mbox{D-MIMO} based on \gls{aoa}, 
but require phase synchronization between anchors, which may not be available in many practical systems.  

An alternative strategy to model-based positioning is machine learning-based fingerprinting 
\cite{gonultacs2021csi, de2020csi}. However, such an approach depends on labeled data, 
which requires expensive measurement campaigns for every deployment. 

Phase synchronization, as well as deployment-specific measurement campaigns, can be avoided by model-based positioning leveraging the delay-Doppler characteristics~\cite{Shames2013}.  
Delay-Doppler-based positioning is used, for example, in global navigation satellite systems (GNSS), 
where the conventional strategy is a \textit{two-step} approach that first estimates delays and radial velocities, 
and then maps these estimates to the position domain \cite{vincent20delayDoppler2Step}. 
However, it has been shown \cite{Closas07DirectDelayDoppler, Closas17Magazine,LiaLeiMey:Asilomar2023} that a \textit{direct} approach, 
where the position is estimated directly from the receive signal rather than from intermediary delay
and velocity estimates, is more robust. References \cite{Weiss14DelayDopplerPF} and \cite{luo2023towards} have also combined direct delay-Doppler tracking based on particle filters, similar to ours. 
However, since the focus of these works is not on simultaneous positioning and communications, 
their position estimates are formed on the basis of specialized positioning signals. In contrast, 
our method is based on the CSI estimates of conventional OFDM systems. Moreover, these other works evaluate their methods
through simulations that do not contain scattering or blockage,
which can severely impair integrated positioning and communication.
In contrast, we show the efficacy of our method on measured data from a real-world environment that contains both scattering and blockage.

{\slshape Notation:}
Column vectors and matrices are denoted by boldface lowercase (e.g., $\bm{x}$) and uppercase letters (e.g., $\bm{X}$), respectively. 
We use $\bm{x}^\trp$ and $\bm{x}^\herm$ to denote the transpose and Hermitian transpose of $\bm{x}$, respectively. 
The Euclidean norm of $\bm{x}$ is $\lVert \bm{x} \rVert$. 
The $N\times N$ identity matrix is $\bm{I}$, where the size is left implicit. 
The operator $\vectorize: \opC^{M\times N}\to\opC^{MN}: \bm{X}\mapsto \bm{x}$ vectorizes the matrix $\bm{X}$
by stacking its columns on top of each other. 
We use $\exp(\bm{x})$ to denote the vector obtained by applying the exponential function to each entry of $\bm{x}$.

\section{Measurement Scenario}
\begin{figure}[tp]
\centering
\setlength{\figurewidth}{0.90\columnwidth}
\pgfdeclarelayer{background}
\pgfdeclarelayer{foreground}
\pgfsetlayers{background,main,foreground}
\begin{tikzpicture}[spy using outlines={circle, magnification=4, size=1.5cm, connect spies}]
 	 \node[anchor=south west,inner sep=0] (image) at (0,0) {\includegraphics[width=0.99\columnwidth]{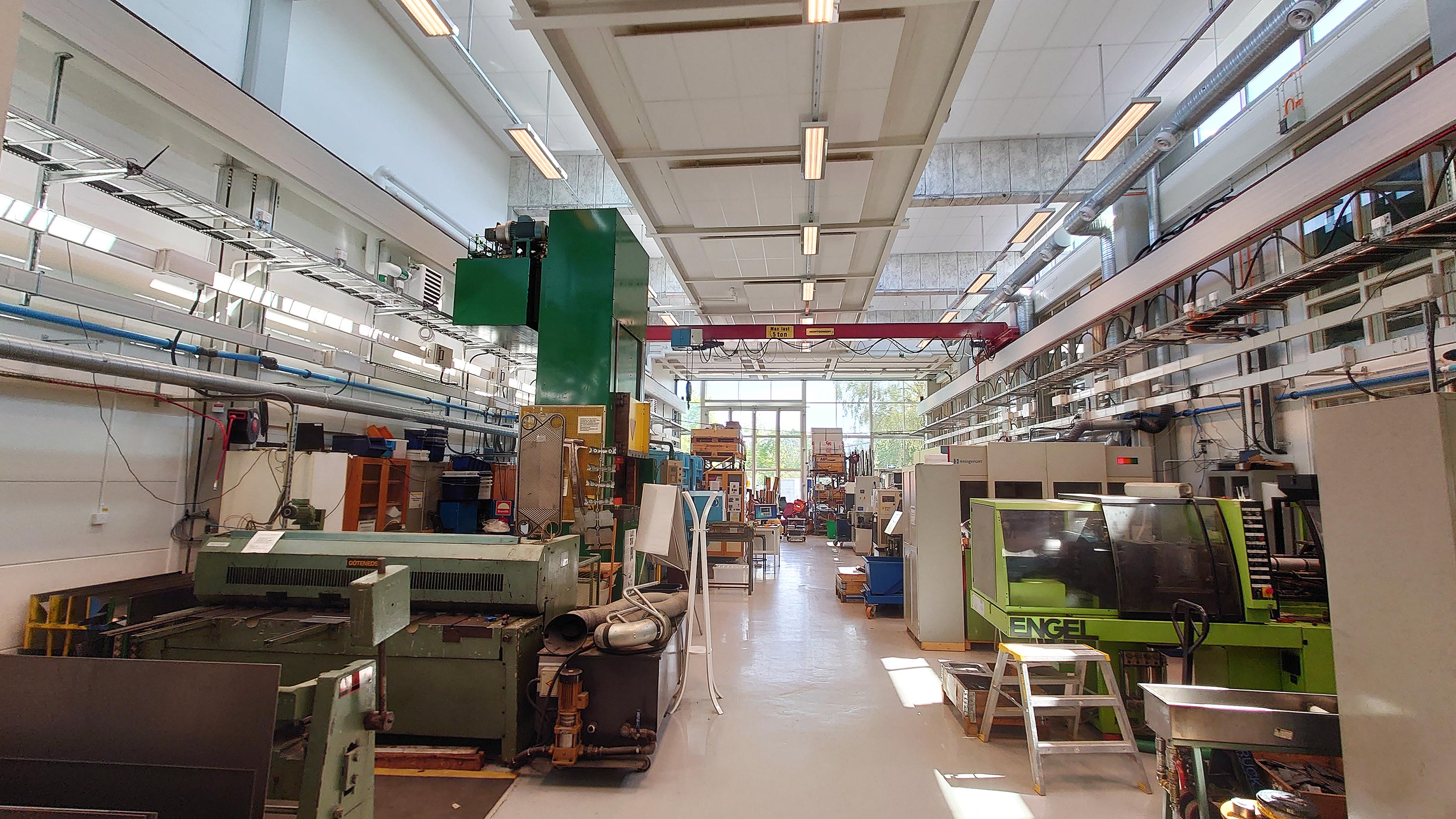}};
	  \spy[myred,ultra thick] on (0.88,4.0) in node at (1.3,2.4);
	  \draw[myred,line width = 0.6pt] (0.88,4.0) circle[radius=0.19cm];
	\draw[myred,line width = 0.6pt] (2.62,3.12) circle[radius=0.1cm];
	\draw[myred,line width = 0.6pt] (3.93,2.49) circle[radius=0.05cm];
	\draw[myred,line width = 0.6pt] (4.07,2.42) circle[radius=0.05cm];
	\draw[myred,line width = 0.6pt] (5.56,2.38) circle[radius=0.05cm];
	\draw[myred,line width = 0.6pt] (5.70,2.44) circle[radius=0.05cm];
	\draw[myred,line width = 0.6pt] (5.89,2.53) circle[radius=0.05cm];
	\draw[myred,line width = 0.6pt] (6.21,2.67) circle[radius=0.05cm];
	\draw[myred,line width = 0.6pt] (6.8,2.94) circle[radius=0.1cm];
	\draw[myred,line width = 0.6pt] (8.15,3.53) circle[radius=0.19cm];
  \begin{pgfonlayer}{foreground}
  	\node[above,align=center] at (1.3,2.48) {\footnotesize anchor\\[-5pt] \footnotesize antenna};
  \end{pgfonlayer}
\end{tikzpicture}
  \caption{Photo of the positioning environment. The industrial hall is approximately \SI{30}{\metre} long and \SI{12}{\metre} wide, with a ceiling height varying between \SI{8}{\metre} and~\SI{12}{\metre} in the center aisle, depending on position.\label{fig:industry-photo}}
\end{figure}
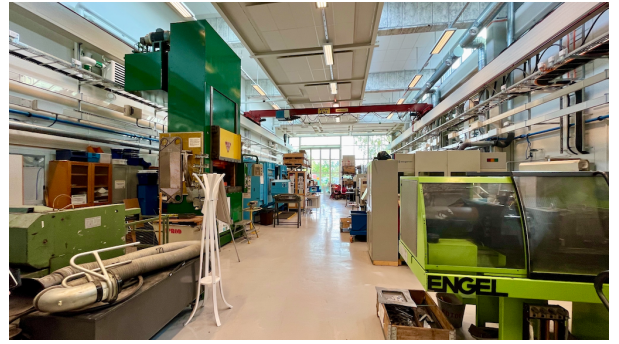

The measurements used for positioning were performed in an industrial environment at the Department of Mechanical Engineering
Sciences at Lund University (see Fig.\;\ref{fig:industry-photo}) with a channel sounder developed at Lund University~\cite{Nelson2024}.
The sounder is designed for D-MIMO and implements an \gls{ofdm} sounding principle.
The carrier frequency was set to $\fcarrier = \SI{3.75}{\giga\hertz}$, with $\Nfrequency = \num{449}$ subcarriers spaced \SI{78.125}{\kilo\hertz} apart, resulting in a measurement bandwidth of $B = \SI{35}{\mega\hertz}$.
All radio units were disciplined with an external \SI{10}{\mega\hertz} clock and a \gls{pps} signal for frequency synchronization, 
but were not calibrated on site and had no phase synchronization. The post-calibrations performed are detailed in~\cite{Nelson2024}. 
Serving as anchors, $M = \num{12}$ single dipole antennas were distributed along the long sides of the industrial hall, at a height of \SI{4}{\metre} above the floor, as indicated in
Fig.\,\ref{fig:scenarios} and \ref{fig:industry-photo}.
To maximize coverage at floor level and mitigate the strong reflection of the walls behind them, 
the antennas were tilted $45^\circ$ downward relative to the wall.
Serving as agent, a remote-controlled robot with a single antenna at \SI{1.35}{\metre} height
was driving through the environment at a maximum speed of \SI{1}{\metre\per\second}.
The agent transmits pilots every $\Delta t \!=\! \SI{5}{\milli\second}$, resulting in 
$200$ channel estimates per second.

In this paper we focus on the representative scenarios shown in Fig.\;\ref{fig:scenarios}.
In $\mathrm{LoS}_1$, the agent moves through the middle aisle of the industrial hall and back. 
There are multiple anchor links under LoS conditions throughout the track,  
but machinery blocks many anchors through significant portions.
We refer to the conditions of $\mathrm{LoS}_1$ as ``partial LoS.''
Track $\mathrm{LoS}_2$ is approximately identical to $\mathrm{LoS}_1$ but was recorded at a later time for control purposes. 
In $\mathrm{OLoS}_1$, the agent moves in circles around a piece of
machinery such that there are parts of the trajectory where all anchor links are under \gls{olos} conditions. 
We therefore refer to the conditions of $\mathrm{OLoS}_1$ as ``full OLoS.''

\section{System Model}
\subsection{Channel Model}
We assume that at each sample index $k\in\{1,\dots,K\}$ (corresponding to sample times $t_{\scriptscriptstyle k}\in\{t_{\scriptscriptstyle1},\dots,t_{\scriptscriptstyle K}\}$ 
with $\Delta t\triangleq t_{\scriptscriptstyle k+1}-t_{\scriptscriptstyle k}$), 
every anchor $m\in\{1,\dots,M\}$ obtains a noisy estimate of the channel between itself and the moving~agent
\begin{align}\label{eq:signal-model}
    \channelVecNoisy{m}{k} = \channelVec{m}{k} + \noiseVec{m}{k}.
\end{align}
Here, $\channelVecNoisy{m}{k}\!\in\!\opC^{\Nfrequency}$ is the estimate of the frequency-domain 
channel vector $\channelVec{m}{k}\!\in\!\opC^{\Nfrequency}$ with carrier frequency $\fcarrier$ and baseband frequencies 
$\bm{f}\!\in\!\mathbb{R}^{\Nfrequency}$, and $\noiseVec{m}{k}$ is observation noise
which we model as white Gaussian noise, $\noiseVec{m}{k}\!\!\sim\mathcal{CN}(\bm{0},\sigma^2\bm{I})$.
For simplicity, we do not explicitly model scattering and blockage due to the machinery and large amount of metal surfaces in the environment (see Fig.\,\ref{fig:industry-photo}). 
Instead, we model the channels as pure \gls{los} links.
We assume that the channel amplitudes remain approximately constant for $\Ntime$ subsequent samples.
We denote the unknown position and (assumed planar) 
velocity of the moving agent at time index~$k$ by
$\pos{k}\in\mathbb{R}^3$ and $\vel{k}=\big[\velComponent{k}{x},\velComponent{k}{y},\velComponent{k}{z}\big]^\trp\!\in\mathbb{R}^2\times\{0\}$, respectively, 
and we assume that 
the velocity remains approximately constant for $\Ntime$ subsequent samples.
The position of the $m$th anchor is denoted by $\posAnchor{m}\in\mathbb{R}^3$, and 
is assumed to be~known. 
These assumptions allow us to model $\Ntime$ subsequent noise-free channel vectors 
$\NoiseFreeChannelMatrix{m}{k} = [\channelVec{m}{k-\Ntime+1}, \dots, \channelVec{m}{k}] \in \complexset{\Nfrequency}{\Ntime}$ as\footnote{For ease of notation, we assume here that estimates of $\Ntime$ channel vectors 
have already been acquired at time step $k=1$.}
\vspace{-0.5mm}
\begin{align}
    \NoiseFreeChannelMatrix{m}{k} &= \amplitude{m}{k} \gresponseTemporal{m}{k} {\gresponseDoppler{m}{k}}^\trp, \label{eq:array_responses}
    \\[-6mm]
       \nonumber
\end{align}
where $\amplitude{m}{k}\!\in\!\opC$ is the complex channel amplitude between the moving agent 
and the $m$th anchor at sample index $k$, \mbox{$\gresponseTemporal{m}{k}\!\in\! \opC^{\Nfrequency}$} is the delay response vector, 
and $\gresponseDoppler{m}{k}\!\in\!\opC^{\Ntime}$ is the Doppler response vector.

The Doppler response vector $\gresponseDoppler{m}{k}$ 
captures how the channel phases evolve during the time frame $\bm{t}=[t_{\scriptscriptstyle k-\Ntime+1},\dots,t_{\scriptscriptstyle k}]^\trp$ due to the velocity of the moving agent $\vel{k}$, and is written as
\begin{align}\label{eq:DopplerResponse}
    \gresponseDoppler{m}{k} =
    \exp\bigg(\mathrm{j}2\pi \frac{\fcarrier}{\mathsf{c}} \bigg\langle\! \frac{\posAnchor{m}-\pos{k}}{\|\posAnchor{m}-\pos{k}\|},\vel{k}\!\bigg\rangle\, \bm{t} \bigg) ,
\end{align}
where $\mathsf{c}$ is the speed of light, and where the inner product represents the radial velocity of the moving agent 
into the direction of the $m$th anchor.
The delay response vector $\gresponseTemporal{m}{k}$ in \eqref{eq:array_responses} 
accounts for the different phase shifts at the different frequencies $\bm{f}$
due to the propagation delay $\mathsf{c}^{-1}\|\posAnchor{m}-\pos{k}\|$, 
and can be expressed as
\begin{align}\label{eq:DelayResponse}
    \gresponseTemporal{m}{k} = \exp\bigg(-\mathrm{j}2\pi \frac{\|\posAnchor{m}-\pos{k}\|}{\mathsf{c}}\bm{f} \bigg).
\end{align}
Finally, the complex amplitude $\amplitude{m}{k}$ in \eqref{eq:array_responses} accounts for path loss, 
the common phase shift at the carrier frequency~$\fcarrier$ due to propagation delay $\mathsf{c}^{-1}\|\posAnchor{m}-\pos{k}\|$, and phase offset at the $m$th anchor due to the absence of phase synchronization. 

Our goal is to estimate the agent trajectory $\{\pos{k}\}_{k=1}^{K}$ based on \gls{dmimo} observations 
    $\meas{k} = \left[\responseMeas{1}{k}, \dots, \responseMeas{M}{k}\right]
    \in\opC^{\Ntime\Nfrequency\times M}$
where
$\responseMeas{m}{k} := \vectorize(\responseMeasMatrix{m}{k})$ are vectorized noisy channel measurements $\responseMeasMatrix{m}{k} = [\channelVecNoisy{m}{k-\Ntime+1}, \dots, \channelVecNoisy{m}{k}]$ corresponding to~\eqref{eq:signal-model}.

\subsection{State-Space Model}
The state transition \gls{pdf} $p(\state{k}|\state{k-1})$ of the agent state\footnote{Note that the state vector does not represent the vertical agent velocity $\velComponent{k}{z}$, which is assumed to be zero, nor the complex amplitude $\amplitude{m}{k}$, which we treat as a nuisance parameter, see Sec.\,\ref{sec:likelihood-model}. Note also that the channel noise variance $\sigma^2$ from \eqref{eq:signal-model} is modeled as a (constant) variable of interest.} $\state{k} = [\pos{k}^\trp, \velComponent{k}{x}\!, \velComponent{k}{y}\!, \sigma^2]^\trp \in \setS$  with $\stateSpace = \mathbb{R}^5 \times \mathbb{R}_{\geq0}$ is chosen to be a nearly constant velocity motion model, i.e., 
\begin{align}\label{eq:state-space-model}
    \state{k} = \transitionMatrix \,\state{k-1} + \processNoise{k}
\end{align}
with transition matrix $\transitionMatrix$ given by
\begin{align}
       \transitionMatrix = \begin{bmatrix}
        1 & 0 & 0 & \Delta t & 0 &  0   \\
        0 & 1 & 0 & 0 & \Delta t & 0    \\
        0 & 0 & 1 & 0 & 0 & 0    \\
        0 & 0 & 0 & 1 & 0 & 0    \\
        0 & 0 & 0 & 0 & 1 & 0    \\
        0 & 0 & 0 & 0 & 0 & 1   \\
       \end{bmatrix}
\end{align}
and $\processNoise{k}\sim\mathcal{N}\big(\bm{0},\diag(\sigma_\mathrm{p}^2, \sigma_\mathrm{p}^2, \sigma_\mathrm{h}^2, \sigma_\mathrm{v}^2, \sigma_\mathrm{v}^2, \sigma_\mathrm{S}^2 )\big)$ 
models process noise.

\subsection{Likelihood Model}\label{sec:likelihood-model} 
The dependency of the observations $\meas{k}$ on the states~$\state{k}$ is modeled as follows:
We assume the measurements at the anchors to be independent, so the likelihood function of 
\mbox{$\meas{k}$}
conditional on $\state{k}$ \mbox{and $\amplitudeVec{k} := [\amplitude{1}{k}, \dots, \amplitude{M}{k}]^\trp$} is
\begin{align}\label{eq:likelihood}
    &p(\meas{k}|\state{k},\amplitudeVec{k}) = \frac{\prod\limits_{m=1}^M \exp 
    \left(
        -\frac{1}{\sigma^2} \lVert \responseMeas{m}{k} - \amplitude{m}{k} \responseModel{m}{k}  \rVert^2
    \right) }{\left(\pi \sigma^2\right)^{\Nantennas \Ntime \Nfrequency}} \,,
\end{align}
where $\responseModel{m}{k} \triangleq \vectorize(\gresponseTemporal{m}{k}{\gresponseDoppler{m}{k}}^\trp)$
is the \emph{delay-Doppler response}. 
We treat the unknown amplitudes $\amplitude{m}{k}$ as nuisance parameters.
Hence, we compute the \textit{profile} likelihood function, i.e., we \textit{concentrate} with respect to $\amplitudeVec{k}$ by computing \gls{ml} estimates
conditional on $\state{k}$ (since the delay-Doppler response $\responseModel{m}{k}$ is a function of $\state{k}$)~\cite{Stoica1995ConcentratedLHF}
\begin{align}\label{eq:amplitudeHat}
    \amplitudeHat{m}{k}|\state{k} = \argmax_{\amplitude{m}{k}} p(\responseMeas{m}{k}|\state{k},\amplitude{m}{k}) = \frac{{\responseModel{m}{k}}^\herm \responseMeas{m}{k}}{\lVert \responseModel{m}{k} \rVert^2}.
\end{align}
Reinserting $\amplitudeHat{m}{k}|\state{k}$ in~\eqref{eq:likelihood}, our profile likelihood function yields 
\begin{align}
    p(\meas{k}|\state{k}) = \frac{ \exp \left( -\frac{1}{\sigma^2} \sum\limits_{m=1}^M
   \left\|\OrthogonalProjector{m}{k} \responseMeas{m}{k} \right\|^2 \right) }{\left(\pi \sigma^2\right)^{\Nantennas \Ntime \Nfrequency}} 
 \label{eq:likelihood-concentrated}
\end{align}
where $\OrthogonalProjector{m}{k} = \bm{I} - \frac{\responseModel{m}{k}{\responseModel{m}{k}}^\herm}{\|\responseModel{m}{k}\|^2}$ is the projector onto the~orthogonal complement of the subspace spanned by $\responseModel{m}{k}$.
Note that the profile likelihood function in \eqref{eq:likelihood-concentrated} is a function of the agent's position $\pos{k}$ and velocity $\vel{k}$ as $\OrthogonalProjector{m}{k}$ is a function of the delay-Doppler response~$\responseModel{m}{k}$.

\section{Delay-Doppler Positioning and Tracking}

\subsection{Recursive Bayesian Filtering}\label{sec:BayesianFiltering}
We formulate the tracking problem through the Bayesian filtering equation~\cite{Gustafsson10PFtheory}
\begin{align}\label{eq:posterior}
    p(\state{k}|\meas{1:k}) = \frac{p(\meas{k}|\state{k}) p(\state{k}| \meas{1:k-1})}{p(\meas{k}| \meas{1:k-1})}
\end{align}
that describes the posterior PDF $p(\state{k}|\meas{1:k})$ of the state vector $\state{k}$ at time step $k$ given past observations (i.e.,\,measurements) $\meas{1:k-1}$ and the current observation $\meas{k}$.

The Chapman-Kolmogorov equation relates the prediction PDF $p(\state{k}| \meas{1:k-1})$ in \eqref{eq:posterior}
to the old posterior PDF $p(\state{k-1}| \meas{1:k-1}\!)$ and  
$p(\state{k}|\state{k-1},\meas{1:k-1})$ 
which\,---\,given the first-order Markovity of our model, see \eqref{eq:state-space-model} and \eqref{eq:likelihood}\,---\,equals
the state transition PDF $p(\state{k}|\state{k-1})$, i.e.,
$p(\state{k}|\state{k-1},\meas{1:k-1}) = p(\state{k}|\state{k-1})$.
By the Chapman-Kolmogorov equation, we thus~have 
\begin{align}\label{eq:pred-density}
    p(\state{k}| \meas{1:k-1}) = \int_{\stateSpace} p(\state{k}|\state{k-1}) p(\state{k-1}| \meas{1:k-1}) \, \diff \state{k-1}.
\end{align}

By the law of total probability, the normalization constant in~\eqref{eq:posterior} can be expressed in continuous integral form through 
\begin{align}\label{eq:normalization-constant}
p(\meas{k}| \meas{1:k-1}) = \int_{\stateSpace} p(\meas{k}|\state{k},\meas{1:k-1})p(\state{k}|\meas{1:k-1}) \diff \state{k}.\!
\end{align}

\subsection{Particle-Based Approximation}
We implement the recursive Bayesian filter from Sec.\,\ref{sec:BayesianFiltering} through a particle filter \cite{Gustafsson10PFtheory} that approximates the posterior PDF with a set of $N$ particles with values $\big\{\particle{k}{i}\big\}_{i=1}^N$ and weights $\big\{\weight{k|k}{i}\big\}_{i=1}^N$.
Since we cannot sample efficiently from the posterior PDF, we sample $\particle{k}{i}$ from a proposal PDF $q(\state{k}|\particle{k-1}{i},\meas{k})$ in each step $k$ that we choose to be defined through our state-space model in~\eqref{eq:state-space-model}, i.e., it equals the state transition PDF $p(\state{k}|\particle{k-1}{i})$. 
For this choice, the prediction (i.e., prior) weights are 
\begin{align}\label{eq:prediction-weights}
    \weight{k|k-1}{i} = 
    \frac{p(\particle{k}{i}|\particle{k-1}{i})}{q(\particle{k}{i}|\particle{k-1}{i},\meas{k})} \weight{k-1|k-1}{i} = \weight{k-1|k-1}{i} \, .
\end{align}
The prediction PDF in~\eqref{eq:pred-density} is approximated by
\begin{align}
    \hat{p}(\state{k}|\meas{1:k-1}) = \sum\nolimits_{i=1}^{N} \weight{k|k-1}{i} \delta\!\left(\state{k} - \particle{k}{i}\right)
\end{align}
which leads to the approximation of the posterior PDF 
\begin{align} \label{eq:posterior-approx}
    \hat{p}(\state{k}|\meas{1:k}) &= \sum\limits_{i=1}^{N} \underbrace{c_k^{\scriptscriptstyle-1} p\left(\meas{k}|\particle{k}{i}\right) \weight{k|k-1}{i}}_{\triangleq\weight{k|k}{i} } \delta\!\left(\state{k} - \particle{k}{i}\right),
\end{align}
where $c_k = \sum_{i=1}^N p\!\left(\meas{k}|\particle{k}{i}\right) \weight{k|k-1}{i}$, 
which approximates the normalization constant $p(\meas{k}| \meas{1:k-1})$ in~\eqref{eq:normalization-constant}, 
ensuring that $\sum_{i=1}^{N} \weight{k|k}{i} = 1$.

We estimate the state $\bm{x}_k$ by approximating the \gls{mmse} estimate 
$\state{k}^{\text{\tiny MMSE}}=\mathbb{E}\left(\state{k}|\meas{1:k}\right)$ as
\begin{align}\label{eq:state-estimate}
    \stateEstimate{k} = \int_{\stateSpace}\!\state{k} \hat{p}(\state{k}|\meas{1:k}) \,  \diff \state{k}= \sum\limits_{i=1}^{N}  \particle{k}{i} \weight{k|k}{i} \, ,
\end{align}
and we approximate the corresponding state covariance matrix 
\begin{align}
    \bm{P}_{k}^{\text{\tiny MMSE}} &= 
      \int_{\stateSpace} 
        \left(
        \state{k} - \state{k}^{\text{\tiny MMSE}}
        \right) 
        \left(
        \state{k} - \state{k}^{\text{\tiny MMSE}}
        \right)^\trp \nonumber
      p(\state{k}|\meas{1:k}) \,  \diff \state{k}
\end{align}
as
\begin{align}\label{eq:empiricalCovariance}
    \covarianceEstimate{k} =  \sum\nolimits_{i=1}^{N}
    \left(\particle{k}{i} - \stateEstimate{k}\right)\left(\particle{k}{i} - \stateEstimate{k}\right)^\trp \weight{k|k}{i} \, .
\end{align}

\subsection{Particle Filter Implementation}
Algorithm~\ref{alg:RPF} summarizes the implemented particle filter. 
We initialize $N$ particles by sampling from a uniform PDF 
$\mathcal{U}(\bm{x}_{\text{\tiny min}},\bm{x}_{\text{\tiny max}})$ 
and we assign weights $\weight{0|0}{i} = 1/N$.
The particle-based posterior PDF in~\eqref{eq:posterior-approx} is computed by the loop starting at line~\ref{line:parfor}. 
Although computationally expensive, it is well suited for parallel computing, while the resampling in line~\ref{line:resampling} is generally not parallelizable.
We choose a regularized particle filter~\cite{Musso2001PF} using a Gaussian kernel with an implementation similar to~\cite[Alg.\,6]{Arulampalam02PFtutorial}. 
That is, in each step $k$, we evaluate~\eqref{eq:empiricalCovariance} and decompose $\hat{\bm{P}}_{k} = \bm{L}_k \bm{L}_k^\trp$ using the Cholesky decomposition.
In line~\ref{line:resampling}, we employ systematic resampling
~\cite[Alg.\,2]{Arulampalam02PFtutorial} 
which reduces particle degeneracy and implies equal weights after resampling (see line~\ref{line:resampled-weights}).
After resampling, each particle is convolved (see line~\ref{line:kernelConvolution}) with a Gaussian regularization kernel $K(\state{k})$ with covariance matrix $\covarianceEstimate{k}$ and scaled by the optimal kernel bandwidth $h_{\text{\tiny opt}}$~\cite[p.\,253]{Musso2001PF}, where $\bm{\epsilon}_i\!\sim\!\mathcal{N}(\bm{0},\bm{I})$, 
to counteract particle impoverishment.

\begin{algorithm}[t]
	\LinesNumbered
	\caption{Regularized Particle Filter}\label{alg:RPF}
	\SetKwInOut{Input}{Input}\SetKwInOut{Output}{Output}
	\Input{$N,\bm{x}_{\text{\tiny min}},\bm{x}_{\text{\tiny max}},\{\meas{k}\}_{k=1}^K$} 
	\Output{$\{\stateEstimate{k}\}_{k=1}^K$}	
	$\particle{0}{i} \sim \mathcal{U}(\bm{x}_{\text{\tiny min}},\bm{x}_{\text{\tiny max}})$ and $\weight{0|0}{i} \gets 1/N \ ~ \forall i \in \{1 \dots N\}$\; 
	  \For{$k \gets 1$ \KwTo $K$ \KwBy $1$}{ 
	  	  \For{$i \gets 1$ \KwTo $N$ \KwBy $1$\label{line:parfor}}{ 
            $\particle{k}{i} \sim p(\state{k}|\particle{k-1}{i})$; \hfill \textcolor{gray}{//see \eqref{eq:state-space-model}\hspace{-4mm}}\\
            $\weight{k|k-1}{i} \gets \weight{k-1|k-1}{i}$;\label{line:importance-weight-compensation} \hfill \textcolor{gray}{//see \eqref{eq:prediction-weights}\hspace{-4mm}}\\
            $\widetilde{w}_{\scriptscriptstyle k|k}^{\scriptscriptstyle (i)} \gets p\!\left(\meas{k}|\particle{k}{i}\right) \, \weight{k|k-1}{i}$; \hfill \textcolor{gray}{//see \eqref{eq:likelihood-concentrated}\textsuperscript{$ a$}\hspace{-4mm}}
		}
        $\left\{\weight{k|k}{i} \right\}_{i=1}^N \gets \left\{ \widetilde{w}_{\scriptscriptstyle k|k}^{\scriptscriptstyle (i)}\big/\big(\sum_{i=1}^N \widetilde{w}_{\scriptscriptstyle k|k}^{\scriptscriptstyle (i)}\big) \right\}_{i=1}^N$\;\vspace{0.05cm}
	    $\stateEstimate{k} \gets \sum_{i=1}^{N}  \particle{k}{i} \weight{k|k}{i}$; \hfill \textcolor{gray}{//see \eqref{eq:state-estimate}\hspace{-4mm}}\\
        $\covarianceEstimate{k} \gets \sum\limits_{i=1}^{N} \left(\particle{k}{i} - \stateEstimate{k}\right)\left(\particle{k}{i} - \stateEstimate{k}\right)^\trp \weight{k|k}{i}$; \hfill \textcolor{gray}{//see \eqref{eq:empiricalCovariance}\hspace{-4mm}}\\
        $\particle{k}{i} \gets \texttt{\small resample}(\{\particle{k}{i},\weight{k|k}{i}\})$; \hfill \textcolor{gray}{//see \cite[Alg.\,2]{Arulampalam02PFtutorial}\hspace{-4mm}}\label{line:resampling} \\
        $\{\weight{k|k}{i}\}_{i=1}^{N} \gets 1/N$;\hfill \textcolor{gray}{//due to resampling\hspace{-4mm}}\label{line:resampled-weights}\\
        $\bm{L}_k \gets \texttt{\small cholesky}(\covarianceEstimate{k})$; \hfill \textcolor{gray}{//s.t. $\bm{L}_k \bm{L}_k^\trp=\covarianceEstimate{k}$\hspace{-4mm}}\\
        \For{$i \gets 1$ \KwTo $N$ \KwBy $1$}{ 
            $\bm{\epsilon}_i \sim \mathcal{N}(\bm{0},\bm{I})$\;
            $\particle{k}{i} \gets \particle{k}{i} + h_{\text{\tiny opt}} \bm{L}_k \bm{\epsilon}_i$\label{line:kernelConvolution}\;
        }
    	}
     \vspace{-0.1cm}
\algorithmfootnote{\textsuperscript{$a$}\footnotesize\,Using the Bartlett spectrum from~\cite{Nelson2024} instead of the likelihood function in~\eqref{eq:likelihood-concentrated} during convergence can increase robustness.}
\end{algorithm} \vspace{-0.3cm}
\renewcommand{\trajectoryLW}{1pt}
\newcommand{\trajectoryEstimateLW}{1pt}
\newcommand{\trajectoryEstimateCOL}{mygreen}

\section{Results}
\begin{figure}[tp]
\centering
\setlength{\figurewidth}{0.99\columnwidth}
  \centering
  \includegraphics[trim=2 2 20 0,clip,width=0.99\columnwidth]{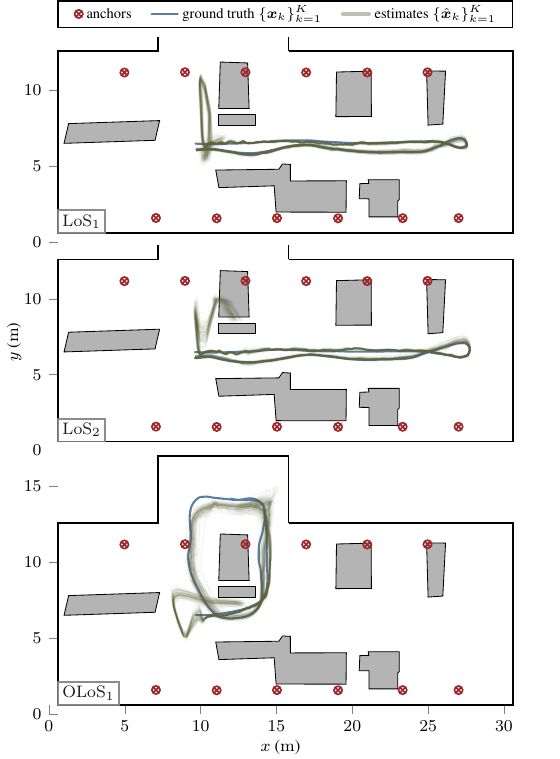}
  \vspace{-0.6cm}\caption{MC analysis showing \num{100} realizations of estimated trajectories vs. true trajectories for each of the tracks $\mathrm{LoS}_1$, $\mathrm{LoS}_2$, and $\mathrm{OLoS}_1$.}
\label{fig:results}
\end{figure}

Algorithm~\ref{alg:RPF} is initialized by drawing $N=16000$ particles\footnote{A smaller number of particles, such as $N=500$, can be chosen to trade off estimation accuracy for faster computation time, which may be suitable for real-time implementations with limited computational resources.} from $\mathcal{U}(\bm{x}_{\text{\tiny min}},\bm{x}_{\text{\tiny max}})$ and is given the channel measurements $\{\meas{k}\}_{k=1}^K$ in each step $k$.
We perform a \gls{mc} analysis with \num{100} runs of each track using different realizations of random numbers (i.e., using different seeds).
Fig.\,\ref{fig:results} shows the estimated trajectories $\{\stateEstimate{k}\}_{k=1}^K$ (\customlineref{color=\trajectoryEstimateCOL, line width=2*\trajectoryEstimateLW, line join=round, line cap=round,draw opacity=0.4}) 
for each of the tracks $\mathrm{LoS}_1$ (top), $\mathrm{LoS}_2$ (center), and $\mathrm{OLoS}_1$ (bottom) in comparison to the ground truth 
(\customlineref{color=mylightblue1, line width=\trajectoryLW, line join=round, line cap=round}).
We initialize by drawing particles uniformly from the spatial region between $[\bm{x}_{\text{\tiny min}}]_{\scriptscriptstyle 1:3} = \bm{0}\,\SI{}{\metre}$ and
$[\bm{x}_{\text{\tiny max}}]_{\scriptscriptstyle 1:3} = [30,15,2.5]^\trp \SI{}{\metre}$, covering the scenario.
Our temporal window size is chosen as $\Ntime\!=\!\num{200}$.
For process noise, we intentionally choose a small $\sigma_\mathrm{p}=\SI{0.3}{\milli\metre}$ and $\sigma_\mathrm{h}\!=\!\SI{2}{\centi\metre}$ to promote that particles stay close to the ``true'' mode (i.e., at $\pos{k}$) in the posterior \gls{pdf}, 
$\sigma_\mathrm{v}\!=\!\SI{0.03}{\metre\per\second}$ and $\sigma_\mathrm{S}\!=\!\SI{0.3}{}$. 
In partial \gls{los}, i.e., when at least some anchors (\,\ref{pgf:antennasIntro}\,) are visible from the agent, our algorithm nearly achieves centimeter-level accuracy.
We analyze the error of the planar \gls{mmse} estimates $\posHest{k}\!:=\![\stateEstimate{k}]_{\scriptscriptstyle 1:2}$ w.r.t. the ground truth $\posH{k}\!:=\![\state{k}]_{\scriptscriptstyle 1:2}$. 
After convergence, we achieve a planar position \gls{rmse} of \SI{10.1}{\centi\metre} on track $\mathrm{LoS}_1$ and an \gls{rmse} of \SI{11.6}{\centi\metre} on track $\mathrm{LoS}_2$ over the steps $k$ of all \num{100} realizations.
On the $\mathrm{OLoS}_1$ track, where parts of the track are in complete OLoS, 
our algorithm achieves an \gls{rmse} of \SI{49.3}{\centi\metre}. 
Fig.\,\ref{fig:CDF} shows the cumulative frequency
of the error $\lVert \posHest{k} - \posH{k} \rVert$ of all $K$ steps from all \num{100} realizations.

\section{Conclusions}
We have demonstrated positioning in an industrial environment using a direct positioning method leveraging delay-Doppler characteristics.
The method is evaluated on real-world D-MIMO measured CSI, which has proper frequency synchronization but lacks phase synchronization. Despite limited bandwidth and a sub-$\SI{6}{\giga\hertz}$ carrier, we show near-centimeter accuracy in partial LoS, and decimeter accuracy in full OLoS.

\begin{figure}[t]
\centering
\setlength{\figurewidth}{0.8\columnwidth}
\setlength{\figureheight}{0.25\columnwidth}
\input{figures/results/CDF}
\vspace{-0.2cm}
\caption{Cumulative frequency of the error of the planar \gls{mmse} estimates w.r.t. the ground truth for all time instances $k$ along the three trajectories.}
  \label{fig:CDF}\vspace{-0.2cm}
\end{figure}

\balance
\bibliographystyle{IEEEtran}
\bibliography{bibtex/bib/IEEEabrv, bibtex/bib/ms}

\end{document}

%% file: abbr.tex
\newacronym{2d}{2D}{two-dimensional}
\newacronym{3d}{3D}{three-dimensional}
\newacronym{6g}{6G}{sixth generation}
\newacronym{adcs}{ADCs}{analog-to-digital converters}
\newacronym{aoa}{AoA}{angle-of-arrival}
\newacronym{aod}{AoD}{angle-of-departure}
\newacronym{asic}{ASIC}{application specific integrated circuit}
\newacronym{awgn}{AWGN}{additive white Gaussian noise}
\newacronym{arfh}{ARFH}{auxiliary RF harvester}
\newacronym{bibc}{BiBC}{bistatic backscatter communication}
\newacronym{bf}{BF}{beamformer}
\newacronym{ce}{CE}{channel estimation}
\newacronym{ced}{CED}{cumulative energy demand}
\newacronym{cdf}{CDF}{cumulative distribution function}
\newacronym{cir}{CIR}{channel impulse response}
\newacronym{csi}{CSI}{channel state information}
\newacronym{crlb}{CRLB}{Cram\'er-Rao lower bound}
\newacronym{dc}{DC}{direct current}
\newacronym{dm}{DM}{diffuse multipath}
\newacronym{dmc}{DMC}{diffuse multipath component}
\newacronym{dli}{DLI}{direct link interference}
\newacronym{dpe}{DPE}{direct position estimation}
\newacronym{eh}{EH}{energy harvesting}
\newacronym{eirp}{EIRP}{equivalent isotropically radiated power}
\newacronym{elaa}{ELAA}{extremely large aperture array}
\newacronym{em}{EM}{electromagnetic}
\newacronym{en}{EN}{energy-neutral}     
\newacronym{end}{END}{energy-neutral device}    
\newacronym{e2e}{E2E}{End-to-End}
\newacronym{epu}{EPU}{edge processing unit}
\newacronym{esd}{ESD}{electrostatic discharge}
\newacronym{fa}{FA}{federation anchor}
\newacronym{jcrs}{JCRS}{joint communication and radar sensing }
\newacronym{ic}{IC}{integrated circuit}
\newacronym{iot}{IoT}{Internet of Things}
\newacronym{id}{ID}{information decoding}
\newacronym{isac}{ISAC}{integrated sensing and communication}
\newacronym{ism}{ISM}{industrial, scientific and medical}
\newacronym{kpi}{KPI}{key performance indicator}
\newacronym{lca}{LCA}{life cycle assessment}
\newacronym{lis}{LIS}{large intelligent surface}
\newacronym{lhf}{LHF}{likelihood function}
\newacronym{liion}{Li-ion}{lithium-ion}
\newacronym{los}{LoS}{line-of-sight}
\newacronym{lmo}{LMO}{lithium ion manganese oxide}
\newacronym{mc}{MC}{Monte Carlo}
\newacronym{mcu}{MCU}{microcontroller unit}
\newacronym{mf}{MF}{matched filter}
\newacronym{ml}{ML}{maximum likelihood}
\newacronym{mmse}{MMSE}{minimum mean square error}
\newacronym{mmwave}{mmWave}{millimeter wave}
\newacronym{mobc}{MoBC}{monostatic backscatter communication}
\newacronym{mimo}{MIMO}{multiple-input multiple-output}
\newacronym{mmimo}{mMIMO}{massive MIMO}
\newacronym{miso}{MISO}{multiple-input single-output}
\newacronym{nlos}{NLoS}{non LoS}
\newacronym{xlmimo}{XL-MIMO}{extremely large-scale MIMO}
\newacronym{dmimo}{D-MIMO}{distributed MIMO}
\newacronym{mrc}{MRC}{maximum ratio combining}
\newacronym{mrfh}{MRFH}{main RF harvester}
\newacronym{mrt}{MRT}{maximum ratio transmission}
\newacronym{ofdm}{OFDM}{orthogonal frequency-division multiplexing}
\newacronym{ota}{OTA}{over-the-air}
\newacronym{olos}{OLoS}{obstructed LoS}
\newacronym{nb}{NB}{narrowband}
\newacronym{nr}{NR}{New Radio}
\newacronym{pa}{PA}{power amplifier}
\newacronym{pf}{PF}{particle filter}
\newacronym{pdf}{PDF}{probability density function}
\newacronym{pdp}{PDP}{power delay profile}
\newacronym{per}{PER}{packet error rate}
\newacronym{pl}{PL}{path loss}
\newacronym{pg}{PG}{path gain}
\newacronym{pc}{PC}{pilot count}
\newacronym{pw}{PW}{planar wavefront}
\newacronym{rcs}{RCS}{radar cross-section}
\newacronym{re}{RE}{radio element}
\newacronym{rf}{RF}{radio frequency}
\newacronym{rfid}{RFID}{\acrshort{rf} identification}
\newacronym{ris}{RIS}{reflective intelligent surface}
\newacronym{rms}{RMS}{root mean square}
 \newacronym{rmse}{RMSE}{root mean square error}
\newacronym{rss}{RSS}{received signal strength}
\newacronym{rssi}{RSSI}{received signal strength indicator}
\newacronym{rw}{RW}{RadioWeave}
\newacronym{sa}{SA}{synthetic aperture}
\newacronym{sdg}{SDG}{Sustainable Development Goal}
\newacronym{sdn}{SDN}{software-defined network}
\newacronym{si}{SI}{self-interference}
\newacronym{sinr}{SINR}{signal-to-interference-plus-noise ratio}
\newacronym{siso}{SISO}{single-input single-output}
\newacronym{slam}{SLAM}{simultaneous localization and mapping}
\newacronym{smc}{SMC}{specular multipath component}
\newacronym{snr}{SNR}{signal-to-noise ratio}
\newacronym{s-parameter}{S-parameter}{scattering parameter}
\newacronym{svd}{SVD}{singular value decomposition}
\newacronym{sw}{SW}{spherical wavefront}
\newacronym{swipt}{SWIPT}{simultaneous wireless information and power transfer}
\newacronym{tdd}{TDD}{time division duplexing}
\newacronym{tdoa}{TDOA}{time-difference-of-arrival}
\newacronym{toa}{TOA}{time-of-arrival}
\newacronym{tosm}{TOSM}{through–open–short–match}
\newacronym{ue}{UE}{user equipment}
\newacronym{uhf}{UHF}{ultra high frequency}
\newacronym{ula}{ULA}{uniform linear array}
\newacronym{upa}{UPA}{uniform planar array}
\newacronym{ura}{URA}{uniform rectangular array}
\newacronym{uwb}{UWB}{ultrawideband}
\newacronym{vna}{VNA}{vector network analyzer}
\newacronym{wb}{WB}{wideband}
\newacronym{wpt}{WPT}{wireless power transfer}
\newacronym{wsn}{WSN}{wireless sensor network}
\newacronym{xets}{XETS}{cross exponentially tapered slot}
\newacronym{zf}{ZF}{zero forcing}
\newacronym{csp}{CSP}{contact service point}
\newacronym{ecsp}{ECSP}{edge computing service point}
\newacronym{ap}{AP}{access point}
\newacronym{pps}{1PPS}{1 pulse per second}

%% file: macros/vmr-symbols-vecbold.tex
\usepackage{amssymb}
\usepackage{amsfonts}
\usepackage{mathrsfs}
\usepackage{xspace}
\usepackage{bm}
\usepackage{upgreek}

\newcommand{\safemath}[2]{\newcommand{#1}{\ensuremath{#2}\xspace}}

\safemath{\bma}{\mathbf{a}}
\safemath{\bmb}{\mathbf{b}}
\safemath{\bmc}{\mathbf{c}}
\safemath{\bmd}{\mathbf{d}}
\safemath{\bme}{\mathbf{e}}
\safemath{\bmf}{\mathbf{f}}
\safemath{\bmg}{\mathbf{g}}
\safemath{\bmh}{\mathbf{h}}
\safemath{\bmi}{\mathbf{i}}
\safemath{\bmj}{\mathbf{j}}
\safemath{\bmk}{\mathbf{k}}
\safemath{\bml}{\mathbf{l}}
\safemath{\bmm}{\mathbf{m}}
\safemath{\bmn}{\mathbf{n}}
\safemath{\bmo}{\mathbf{o}}
\safemath{\bmp}{\mathbf{p}}
\safemath{\bmq}{\mathbf{q}}
\safemath{\bmr}{\mathbf{r}}
\safemath{\bms}{\mathbf{s}}
\safemath{\bmt}{\mathbf{t}}
\safemath{\bmu}{\mathbf{u}}
\safemath{\bmv}{\mathbf{v}}
\safemath{\bmw}{\mathbf{w}}
\safemath{\bmx}{\mathbf{x}}
\safemath{\bmy}{\mathbf{y}}
\safemath{\bmz}{\mathbf{z}}
\safemath{\bmzero}{\mathbf{0}}
\safemath{\bmone}{\mathbf{1}}

\bmdefine{\biad}{a}
\bmdefine{\bibd}{b}
\bmdefine{\bicd}{c}
\bmdefine{\bidd}{d}
\bmdefine{\bied}{e}
\bmdefine{\bifd}{f}
\bmdefine{\bigd}{g}
\bmdefine{\bihd}{h}
\bmdefine{\biid}{i}
\bmdefine{\bijd}{j}
\bmdefine{\bikd}{k}
\bmdefine{\bild}{l}
\bmdefine{\bimd}{m}
\bmdefine{\bind}{n}
\bmdefine{\biod}{o}
\bmdefine{\bipd}{p}
\bmdefine{\biqd}{q}
\bmdefine{\bird}{r}
\bmdefine{\bisd}{s}
\bmdefine{\bitd}{t}
\bmdefine{\biud}{u}
\bmdefine{\bivd}{v}
\bmdefine{\biwd}{w}
\bmdefine{\bixd}{x}
\bmdefine{\biyd}{y}
\bmdefine{\bizd}{z}

\bmdefine{\bixid}{\xi}
\bmdefine{\bilambdad}{\lambda}
\bmdefine{\bimud}{\mu}
\bmdefine{\bithetad}{\theta}
\bmdefine{\biphid}{\phi}
\bmdefine{\bideltad}{\delta}

\safemath{\bmia}{\biad}
\safemath{\bmib}{\bibd}
\safemath{\bmic}{\bicd}
\safemath{\bmid}{\bidd}
\safemath{\bmie}{\bied}
\safemath{\bmif}{\bifd}
\safemath{\bmig}{\bigd}
\safemath{\bmih}{\bihd}
\safemath{\bmii}{\biid}
\safemath{\bmij}{\bijd}
\safemath{\bmik}{\bikd}
\safemath{\bmil}{\bild}
\safemath{\bmim}{\bimd}
\safemath{\bmin}{\bind}
\safemath{\bmio}{\biod}
\safemath{\bmip}{\bipd}
\safemath{\bmiq}{\biqd}
\safemath{\bmir}{\bird}
\safemath{\bmis}{\bisd}
\safemath{\bmit}{\bitd}
\safemath{\bmiu}{\biud}
\safemath{\bmiv}{\bivd}
\safemath{\bmiw}{\biwd}
\safemath{\bmix}{\bixd}
\safemath{\bmiy}{\biyd}
\safemath{\bmiz}{\bizd}

\safemath{\bmxi}{\bixid}
\safemath{\bmlambda}{\bilambdad}
\safemath{\bmmu}{\bimud}
\safemath{\bmtheta}{\bithetad}
\safemath{\bmphi}{\biphid}
\safemath{\bmdelta}{\bideltad}

\safemath{\bA}{\mathbf{A}}
\safemath{\bB}{\mathbf{B}}
\safemath{\bC}{\mathbf{C}}
\safemath{\bD}{\mathbf{D}}
\safemath{\bE}{\mathbf{E}}
\safemath{\bF}{\mathbf{F}}
\safemath{\bG}{\mathbf{G}}
\safemath{\bH}{\mathbf{H}}
\safemath{\bI}{\mathbf{I}}
\safemath{\bJ}{\mathbf{J}}
\safemath{\bK}{\mathbf{K}}
\safemath{\bL}{\mathbf{L}}
\safemath{\bM}{\mathbf{M}}
\safemath{\bN}{\mathbf{N}}
\safemath{\bO}{\mathbf{O}}
\safemath{\bP}{\mathbf{P}}
\safemath{\bQ}{\mathbf{Q}}
\safemath{\bR}{\mathbf{R}}
\safemath{\bS}{\mathbf{S}}
\safemath{\bT}{\mathbf{T}}
\safemath{\bU}{\mathbf{U}}
\safemath{\bV}{\mathbf{V}}
\safemath{\bW}{\mathbf{W}}
\safemath{\bX}{\mathbf{X}}
\safemath{\bY}{\mathbf{Y}}
\safemath{\bZ}{\mathbf{Z}}

\safemath{\bZero}{\mathbf{0}}
\safemath{\bOne}{\mathbf{1}}
\safemath{\bDelta}{\mathbf{\Delta}}
\safemath{\bLambda}{\mathbf{\UpLambda}}
\safemath{\bPhi}{\mathbf{\Upphi}}
\safemath{\bSigma}{\mathbf{\Upsigma}}
\safemath{\bOmega}{\mathbf{\Upomega}}
\safemath{\bTheta}{\mathbf{\Uptheta}}

\bmdefine{\biAd}{A}
\bmdefine{\biBd}{B}
\bmdefine{\biCd}{C}
\bmdefine{\biDd}{D}
\bmdefine{\biEd}{E}
\bmdefine{\biFd}{F}
\bmdefine{\biGd}{G}
\bmdefine{\biHd}{H}
\bmdefine{\biId}{I}
\bmdefine{\biJd}{J}
\bmdefine{\biKd}{K}
\bmdefine{\biLd}{L}
\bmdefine{\biMd}{M}
\bmdefine{\biOd}{N}
\bmdefine{\biPd}{O}
\bmdefine{\biQd}{P}
\bmdefine{\biRd}{R}
\bmdefine{\biSd}{S}
\bmdefine{\biTd}{T}
\bmdefine{\biUd}{U}
\bmdefine{\biVd}{V}
\bmdefine{\biWd}{W}
\bmdefine{\biXd}{X}
\bmdefine{\biYd}{Y}
\bmdefine{\biZd}{Z}

\bmdefine{\biDelta}{\Delta}
\bmdefine{\biLambda}{\Lambda}
\bmdefine{\biPhi}{\Phi}
\bmdefine{\biSigma}{\Sigma}
\bmdefine{\biOmega}{\Omega}
\bmdefine{\biTheta}{\Theta}

\safemath{\bimA}{\biAd}
\safemath{\bimB}{\biBd}
\safemath{\bimC}{\biCd}
\safemath{\bimD}{\biDd}
\safemath{\bimE}{\biEd}
\safemath{\bimF}{\biFd}
\safemath{\bimG}{\biGd}
\safemath{\bimH}{\biHd}
\safemath{\bimI}{\biId}
\safemath{\bimJ}{\biJd}
\safemath{\bimK}{\biKd}
\safemath{\bimL}{\biLd}
\safemath{\bimM}{\biMd}
\safemath{\bimN}{\biNd}
\safemath{\bimO}{\biOd}
\safemath{\bimP}{\biPd}
\safemath{\bimQ}{\biQd}
\safemath{\bimR}{\biRd}
\safemath{\bimS}{\biSd}
\safemath{\bimT}{\biTd}
\safemath{\bimU}{\biUd}
\safemath{\bimV}{\biVd}
\safemath{\bimW}{\biWd}
\safemath{\bimX}{\biXd}
\safemath{\bimY}{\biYd}
\safemath{\bimZ}{\biZd}

\safemath{\bimDelta}{\biDelta}
\safemath{\bimLambda}{\biLambda}
\safemath{\bimPhi}{\biPhi}
\safemath{\bimSigma}{\biSigma}
\safemath{\bimOmega}{\biOmega}
\safemath{\bimTheta}{\biTheta}

\safemath{\setA}{\mathcal{A}}
\safemath{\setB}{\mathcal{B}}
\safemath{\setC}{\mathcal{C}}
\safemath{\setD}{\mathcal{D}}
\safemath{\setE}{\mathcal{E}}
\safemath{\setF}{\mathcal{F}}
\safemath{\setG}{\mathcal{G}}
\safemath{\setH}{\mathcal{H}}
\safemath{\setI}{\mathcal{I}}
\safemath{\setJ}{\mathcal{J}}
\safemath{\setK}{\mathcal{K}}
\safemath{\setL}{\mathcal{L}}
\safemath{\setM}{\mathcal{M}}
\safemath{\setN}{\mathcal{N}}
\safemath{\setO}{\mathcal{O}}
\safemath{\setP}{\mathcal{P}}
\safemath{\setQ}{\mathcal{Q}}
\safemath{\setR}{\mathcal{R}}
\safemath{\setS}{\mathcal{S}}
\safemath{\setT}{\mathcal{T}}
\safemath{\setU}{\mathcal{U}}
\safemath{\setV}{\mathcal{V}}
\safemath{\setW}{\mathcal{W}}
\safemath{\setX}{\mathcal{X}}
\safemath{\setY}{\mathcal{Y}}
\safemath{\setZ}{\mathcal{Z}}
\safemath{\emptySet}{\varnothing}

\safemath{\colA}{\mathscr{A}}
\safemath{\colB}{\mathscr{B}}
\safemath{\colC}{\mathscr{C}}
\safemath{\colD}{\mathscr{D}}
\safemath{\colE}{\mathscr{E}}
\safemath{\colF}{\mathscr{F}}
\safemath{\colG}{\mathscr{G}}
\safemath{\colH}{\mathscr{H}}
\safemath{\colI}{\mathscr{I}}
\safemath{\colJ}{\mathscr{J}}
\safemath{\colK}{\mathscr{K}}
\safemath{\colL}{\mathscr{L}}
\safemath{\colM}{\mathscr{M}}
\safemath{\colN}{\mathscr{N}}
\safemath{\colO}{\mathscr{O}}
\safemath{\colP}{\mathscr{P}}
\safemath{\colQ}{\mathscr{Q}}
\safemath{\colR}{\mathscr{R}}
\safemath{\colS}{\mathscr{S}}
\safemath{\colT}{\mathscr{T}}
\safemath{\colU}{\mathscr{U}}
\safemath{\colV}{\mathscr{V}}
\safemath{\colW}{\mathscr{W}}
\safemath{\colX}{\mathscr{X}}
\safemath{\colY}{\mathscr{Y}}
\safemath{\colZ}{\mathscr{Z}}

\safemath{\opA}{\mathbb{A}}
\safemath{\opB}{\mathbb{B}}
\safemath{\opC}{\mathbb{C}}
\safemath{\opD}{\mathbb{D}}
\safemath{\opE}{\mathbb{E}}
\safemath{\opF}{\mathbb{F}}
\safemath{\opG}{\mathbb{G}}
\safemath{\opH}{\mathbb{H}}
\safemath{\opI}{\mathbb{I}}
\safemath{\opJ}{\mathbb{J}}
\safemath{\opK}{\mathbb{K}}
\safemath{\opL}{\mathbb{L}}
\safemath{\opM}{\mathbb{M}}
\safemath{\opN}{\mathbb{N}}
\safemath{\opO}{\mathbb{O}}
\safemath{\opP}{\mathbb{P}}
\safemath{\opQ}{\mathbb{Q}}
\safemath{\opR}{\mathbb{R}}
\safemath{\opS}{\mathbb{S}}
\safemath{\opT}{\mathbb{T}}
\safemath{\opU}{\mathbb{U}}
\safemath{\opV}{\mathbb{V}}
\safemath{\opW}{\mathbb{W}}
\safemath{\opX}{\mathbb{X}}
\safemath{\opY}{\mathbb{Y}}
\safemath{\opZ}{\mathbb{Z}}
\safemath{\opZero}{\mathbb{O}}
\safemath{\identityop}{\opI}

\safemath{\veca}{\bma}
\safemath{\vecb}{\bmb}
\safemath{\vecc}{\bmc}
\safemath{\vecd}{\bmd}
\safemath{\vece}{\bme}
\safemath{\vecf}{\bmf}
\safemath{\vecg}{\bmg}
\safemath{\vech}{\bmh}
\safemath{\veci}{\bmi}
\safemath{\vecj}{\bmj}
\safemath{\veck}{\bmk}
\safemath{\vecl}{\bml}
\safemath{\vecm}{\bmm}
\safemath{\vecn}{\bmn}
\safemath{\veco}{\bmo}
\safemath{\vecp}{\bmp}
\safemath{\vecq}{\bmq}
\safemath{\vecr}{\bmr}
\safemath{\vecs}{\bms}
\safemath{\vect}{\bmt}
\safemath{\vecu}{\bmu}
\safemath{\vecv}{\bmv}
\safemath{\vecw}{\bmw}
\safemath{\vecx}{\bmx}
\safemath{\vecy}{\bmy}
\safemath{\vecz}{\bmz}

\safemath{\veczero}{\bmzero}
\safemath{\vecone}{\bmone}
\safemath{\vecxi}{\bmxi}
\safemath{\veclambda}{\bmlambda}
\safemath{\vecmu}{\bmmu}
\safemath{\vectheta}{\bmtheta}
\safemath{\vecphi}{\bmphi}
\safemath{\vecdelta}{\bmdelta}

\safemath{\matA}{\bA}
\safemath{\matB}{\bB}
\safemath{\matC}{\bC}
\safemath{\matD}{\bD}
\safemath{\matE}{\bE}
\safemath{\matF}{\bF}
\safemath{\matG}{\bG}
\safemath{\matH}{\bH}
\safemath{\matI}{\bI}
\safemath{\matJ}{\bJ}
\safemath{\matK}{\bK}
\safemath{\matL}{\bL}
\safemath{\matM}{\bM}
\safemath{\matN}{\bN}
\safemath{\matO}{\bO}
\safemath{\matP}{\bP}
\safemath{\matQ}{\bQ}
\safemath{\matR}{\bR}
\safemath{\matS}{\bS}
\safemath{\matT}{\bT}
\safemath{\matU}{\bU}
\safemath{\matV}{\bV}
\safemath{\matW}{\bW}
\safemath{\matX}{\bX}
\safemath{\matY}{\bY}
\safemath{\matZ}{\bZ}
\safemath{\matzero}{\bmzero}

\safemath{\matDelta}{\bDelta}
\safemath{\matLambda}{\bLambda}
\safemath{\matPhi}{\bPhi}
\safemath{\matSigma}{\bSigma}
\safemath{\matOmega}{\bOmega}
\safemath{\matTheta}{\bTheta}

\safemath{\matidentity}{\matI}
\safemath{\matone}{\matO}

\safemath{\rnda}{A}
\safemath{\rndb}{B}
\safemath{\rndc}{C}
\safemath{\rndd}{D}
\safemath{\rnde}{E}
\safemath{\rndf}{F}
\safemath{\rndg}{G}
\safemath{\rndh}{H}
\safemath{\rndi}{I}
\safemath{\rndj}{J}
\safemath{\rndk}{K}
\safemath{\rndl}{L}
\safemath{\rndm}{M}
\safemath{\rndn}{N}
\safemath{\rndo}{O}
\safemath{\rndp}{P}
\safemath{\rndq}{Q}
\safemath{\rndr}{R}
\safemath{\rnds}{S}
\safemath{\rndt}{T}
\safemath{\rndu}{U}
\safemath{\rndv}{V}
\safemath{\rndw}{W}
\safemath{\rndx}{X}
\safemath{\rndy}{Y}
\safemath{\rndz}{Z}

\safemath{\rveca}{\bimA}
\safemath{\rvecb}{\bimB}
\safemath{\rvecc}{\bimC}
\safemath{\rvecd}{\bimD}
\safemath{\rvece}{\bimE}
\safemath{\rvecf}{\bimF}
\safemath{\rvecg}{\bimG}
\safemath{\rvech}{\bimH}
\safemath{\rveci}{\bimI}
\safemath{\rvecj}{\bimJ}
\safemath{\rveck}{\bimK}
\safemath{\rvecl}{\bimL}
\safemath{\rvecm}{\bimM}
\safemath{\rvecn}{\bimN}
\safemath{\rveco}{\bomO}
\safemath{\rvecp}{\bimP}
\safemath{\rvecq}{\bimQ}
\safemath{\rvecr}{\bimR}
\safemath{\rvecs}{\bimS}
\safemath{\rvect}{\bimT}
\safemath{\rvecu}{\bimU}
\safemath{\rvecv}{\bimV}
\safemath{\rvecw}{\bimW}
\safemath{\rvecx}{\bimX}
\safemath{\rvecy}{\bimY}
\safemath{\rvecz}{\bimZ}

\safemath{\rvecxi}{\bmxi}
\safemath{\rveclambda}{\bmlambda}
\safemath{\rvecmu}{\bmmu}
\safemath{\rvectheta}{\bmtheta}
\safemath{\rvecphi}{\bmphi}

\safemath{\rmatA}{\bimA}
\safemath{\rmatB}{\bimB}
\safemath{\rmatC}{\bimC}
\safemath{\rmatD}{\bimD}
\safemath{\rmatE}{\bimE}
\safemath{\rmatF}{\bimF}
\safemath{\rmatG}{\bimG}
\safemath{\rmatH}{\bimH}
\safemath{\rmatI}{\bimI}
\safemath{\rmatJ}{\bimJ}
\safemath{\rmatK}{\bimK}
\safemath{\rmatL}{\bimL}
\safemath{\rmatM}{\bimM}
\safemath{\rmatN}{\bimN}
\safemath{\rmatO}{\bimO}
\safemath{\rmatP}{\bimP}
\safemath{\rmatQ}{\bimQ}
\safemath{\rmatR}{\bimR}
\safemath{\rmatS}{\bimS}
\safemath{\rmatT}{\bimT}
\safemath{\rmatU}{\bimU}
\safemath{\rmatV}{\bimV}
\safemath{\rmatW}{\bimW}
\safemath{\rmatX}{\bimX}
\safemath{\rmatY}{\bimY}
\safemath{\rmatZ}{\bimZ}

\safemath{\rmatDelta}{\bimDelta}
\safemath{\rmatLambda}{\bimLambda}
\safemath{\rmatPhi}{\bimPhi}
\safemath{\rmatSigma}{\bimSigma}
\safemath{\rmatOmega}{\bimOmega}
\safemath{\rmatTheta}{\bimTheta}

%% file: math-notations.tex
\newcommand{\onematrix}[2]{\bm{1}_{\scriptsize{#1\times#2}}}        
\newcommand{\diff}[0]{\mathrm{d}}                                   

\newcommand{\Nstate}[0]{\scriptscriptstyle \mathcal{D}}             
\newcommand{\Nfrequency}[0]{N_\text{\tiny f}}                       
\newcommand{\nfrequency}[0]{n_\text{\tiny f}}                       
\newcommand{\Ntime}[0]{N_\text{\tiny t}}                            
\newcommand{\Nantennas}[0]{M}                                       

\newcommand{\fcarrier}[0]{\mathsf{f}_{\scriptscriptstyle \mathsf{c}}}                         
\newcommand{\frep}[0]{f_\text{\tiny rep}}                           

\newcommand{\channelVecNoisy}[2]{\widetilde{\bm{h}}_{\scriptscriptstyle#2}^{\scriptscriptstyle(#1)}}   
\newcommand{\channelVec}[2]{{\bm{h}}_{\scriptscriptstyle#2}^{\scriptscriptstyle(#1)}}   
\newcommand{\noiseVec}[2]{{\bm{n}}_{\scriptscriptstyle#2}^{\scriptscriptstyle(#1)}}   
\newcommand{\amplitude}[2]{{\alpha}_{\scriptscriptstyle#2}^{\scriptscriptstyle(#1)}}   
\newcommand{\amplitudeVec}[1]{{\bm{\alpha}}_{\scriptscriptstyle#1}}   
\newcommand{\amplitudeVecHat}[1]{\hat{\bm{\alpha}}_{\scriptscriptstyle#1}}   
\newcommand{\amplitudeHat}[2]{\hat{\alpha}_{\scriptscriptstyle#2}^{\scriptscriptstyle(#1)}}   

\newcommand{\responseMeas}[2]{\bm{y}_{\scriptscriptstyle#2}^{\scriptscriptstyle(#1)}}    
\newcommand{\responseMeasMatrix}[2]{\widetilde{\bm{H}}_{\scriptscriptstyle#2}^{\scriptscriptstyle(#1)}}    

\newcommand{\NoiseFreeChannelMatrix}[2]{\bm{H}_{\scriptscriptstyle#2}^{\scriptscriptstyle(#1)}}    

\newcommand{\responseSpatial}[1]{\bm{b}_{\scriptscriptstyle#1}}    
\newcommand{\responseTemporal}[1]{\bm{b}_{\scriptscriptstyle#1}}    
\newcommand{\responseDoppler}[1]{\bm{c}_{\scriptscriptstyle#1}}    
\newcommand{\responseModel}[2]{\bm{\psi}_{\scriptscriptstyle#2}^{\scriptscriptstyle(#1)}}    
\newcommand{\responseModelMatrix}[2]{\bm{\Psi}_{\scriptscriptstyle#2}^{\scriptscriptstyle(#1)}}    
\newcommand{\OrthogonalProjector}[2]{\bm{\Pi}^{\perp}_{\scriptscriptstyle#1,\scriptscriptstyle#2}}    

\newcommand{\gresponseTemporal}[2]{\bm{b}_{\scriptscriptstyle#2}^{\scriptscriptstyle(#1)}}     
\newcommand{\gresponseDoppler}[2]{\bm{c}_{\scriptscriptstyle#2}^{\scriptscriptstyle(#1)}}    

\newcommand{\state}[1]{\bm{x}_{\scriptscriptstyle#1}}               
\newcommand{\meas}[1]{\bm{Y}_{\scriptscriptstyle#1}}                
\newcommand{\particle}[2]{\bm{x}_{\scriptscriptstyle #1}^{\scriptscriptstyle(#2)}}     
\newcommand{\weight}[2]{w_{\scriptscriptstyle #1}^{{\scriptscriptstyle(#2)}}}          

\newcommand{\stateSpace}[0]{\mathcal{S}}                            
\newcommand{\transitionMatrix}[0]{\bm{\Phi}}                        
\newcommand{\posAnchor}[1]{\bm{p}^{\scriptscriptstyle(#1)}_{\text{\tiny a}}}  
\newcommand{\pos}[1]{\bm{p}_{\scriptscriptstyle#1}}                                   
\newcommand{\posHest}[1]{\hat{\bm{p}}_{\scriptscriptstyle#1}^{\text{\tiny h}}}                                   
\newcommand{\posH}[1]{\bm{p}_{\scriptscriptstyle#1}^{\text{\tiny h}}}                                   
\newcommand{\range}[2]{\bm{r}_{\scriptscriptstyle#1}^{\scriptscriptstyle(#2)}}                                   
\newcommand{\delay}[2]{\tau_{\scriptscriptstyle#1}^{\scriptscriptstyle(#2)}}                                   
\newcommand{\vel}[1]{\bm{v}_{\scriptscriptstyle#1}}                                   
\newcommand{\velComponent}[2]{{v}_{\scriptscriptstyle#1}^{\scriptscriptstyle(#2)}}                        
\newcommand{\velRadial}[2]{v_{\text{\tiny r,}\scriptscriptstyle#1}^{\scriptscriptstyle(#2)}}                                   
\newcommand{\processNoise}[1]{\bm{w}_{\scriptscriptstyle#1}}        
\newcommand{\processNoiseCovariance}[0]{\bm{Q}} 

\newcommand{\stateEstimate}[1]{\hat{\bm{x}}_{\scriptscriptstyle#1}}         
\newcommand{\covarianceEstimate}[1]{\hat{\bm{P}}_{\scriptscriptstyle#1}}    

%% file: macros/colorsGian.tex
\definecolor{myred}{HTML}{9E292B}
\definecolor{myblue}{HTML}{235787}
\definecolor{mygreen}{HTML}{5E6638}
\definecolor{mygray}{HTML}{444444}
\definecolor{myblack}{HTML}{000000}
\definecolor{mywhite}{HTML}{FFFFFF}

\definecolor{myaltred}{HTML}{D46A78}
\definecolor{myaltblue}{HTML}{6699C2}
\definecolor{myaltgreen}{HTML}{B0B58C}
\definecolor{myaltgray}{HTML}{AAAAAA}

\definecolor{mylightred1}{HTML}{B15455}
\definecolor{mylightred2}{HTML}{C57F80}
\definecolor{mylightred3}{HTML}{D8A9AA}
\definecolor{mylightred4}{HTML}{ECD4D5}
\definecolor{mylightblue1}{HTML}{5A7DA5}
\definecolor{mylightblue2}{HTML}{7D99BA}
\definecolor{mylightblue3}{HTML}{B3C3D7}
\definecolor{mylightblue4}{HTML}{D3DCE8}
\definecolor{mydarkgreen}{HTML}{3E4822}
\definecolor{mylightgreen1}{HTML}{828859}
\definecolor{mylightgreen2}{HTML}{9AA075}
\definecolor{mylightgreen3}{HTML}{B8BC96}
\definecolor{mylightgreen4}{HTML}{D4D4B8}
\definecolor{mylightgray1}{HTML}{6F6F6F}
\definecolor{mylightgray2}{HTML}{999999}
\definecolor{mylightgray3}{HTML}{B4B4B4}
\definecolor{mylightgray4}{HTML}{DCDCDC}

%% file: figures/scenarios_v2.tex

\pgfplotsset{every axis/.append style={
  label style={font=\footnotesize},
  legend style={font=\footnotesize},
  tick label style={font=\footnotesize},
}}

\definecolor{myred}{HTML}{9E292B}
\definecolor{myblue}{HTML}{235787}
\definecolor{mygreen}{HTML}{5E6638}
\definecolor{mygray}{HTML}{444444}
\definecolor{myblack}{HTML}{000000}
\definecolor{mywhite}{HTML}{FFFFFF}

\definecolor{myaltred}{HTML}{D46A78}
\definecolor{myaltblue}{HTML}{6699C2}
\definecolor{myaltgreen}{HTML}{B0B58C}
\definecolor{myaltgray}{HTML}{AAAAAA}

\definecolor{mylightred1}{HTML}{B15455}
\definecolor{mylightred2}{HTML}{C57F80}
\definecolor{mylightred3}{HTML}{D8A9AA}
\definecolor{mylightred4}{HTML}{ECD4D5}
\definecolor{mylightblue1}{HTML}{5A7DA5}
\definecolor{mylightblue2}{HTML}{7D99BA}
\definecolor{mylightblue3}{HTML}{B3C3D7}
\definecolor{mylightblue4}{HTML}{D3DCE8}
\definecolor{mydarkgreen}{HTML}{3E4822}
\definecolor{mylightgreen1}{HTML}{828859}
\definecolor{mylightgreen2}{HTML}{9AA075}
\definecolor{mylightgreen3}{HTML}{B8BC96}
\definecolor{mylightgreen4}{HTML}{D4D4B8}
\definecolor{mylightgray1}{HTML}{6F6F6F}
\definecolor{mylightgray2}{HTML}{999999}
\definecolor{mylightgray3}{HTML}{B4B4B4}
\definecolor{mylightgray4}{HTML}{DCDCDC}


\pgfplotsset{
        legend image code/.code={
            \draw[mark repeat=2,mark phase=2]
            plot coordinates {
                (0cm,0cm)
                (0cm,0cm)        
                (0.3cm,0cm)         
            };%
        }
}

\begin{tikzpicture}

\begin{axis}[%
	axis background/.style={fill=white},
width=\figurewidth,
axis equal image,
at={(0,0)},
scale only axis,
xmin=0,
ymin=0,
tick pos=left,
y label style={yshift=-3mm, xshift=-4mm},
x label style={yshift=1.5mm},
, line cap=round
xmax=31.25,
ymax=17.5,
xlabel={$x$ (\SI{}{\metre})},
ylabel={$y$ (\SI{}{\metre})},
legend columns=3,
legend style={at={(0.9375,0.98)}, anchor=north east, legend cell align=left, align=left, draw=none,fill opacity=0.1,text opacity=1, 
/tikz/column 4/.style={
                column sep=3pt,
            }, 
/tikz/row 1/.style={
                row sep=4pt,
            }
},
every outer y axis line/.append style={opacity=0},
]


\addplot [thick, draw=black, fill=white, forget plot]
table[row sep=crcr]{%
	15.8	12.6\\
	15.8	17\\
	7.2	17\\
	7.2	12.6\\
	0.6	12.6\\
	0.6	0.6\\
	30.6	0.6\\
	30.6	12.6\\
	15.8	12.6\\
};

\addplot[area legend, draw=black, fill=mylightgray3, forget plot]
table[row sep=crcr] {%
	x	y\\
	1	6.5\\
	7	6.7\\
	7.3	8\\
	1.3	7.8\\
}--cycle;

\addplot[thick, area legend, draw=black, fill=mylightgray4, forget plot]
table[row sep=crcr] {%
	x	y\\
	15.8	12.6\\
	15.8	17\\
	7.2	17\\
	7.2	12.6\\
};

\addplot[area legend, draw=black, fill=mylightgray3, forget plot]
table[row sep=crcr] {%
	x	y\\
	21.1	1.65\\
	23	1.65\\
	23	2.7\\
	23.1	2.8\\
	23.1	4.1\\
	21.05	4.1\\
	21.1	3.86\\
	20.5	3.84\\
	20.46	2.85\\
	21.1	2.84\\
}--cycle;

\addplot[area legend, draw=black, fill=mylightgray3, forget plot]
table[row sep=crcr] {%
	x	y\\
	11.2	7.7\\
	11.2	8.4\\
	13.6	8.4\\
	13.6	7.7\\
}--cycle;

\addplot[area legend, draw=black, fill=mylightgray3, forget plot]
table[row sep=crcr] {%
	x	y\\
	11.2	8.8\\
	13.2	8.8\\
	13.1	11.8\\
	11.3	11.86\\
}--cycle;

\addplot[area legend, draw=black, fill=mylightgray3, forget plot]
table[row sep=crcr] {%
	x	y\\
	11.21	3.58\\
	14.85	3.7\\
	14.98	1.97\\
	19.59	1.96\\
	19.61	4.03\\
	15.92	4.02\\
	15.92	5.11\\
	15.4	5.14\\
	15.16	4.79\\
	11	4.74\\
}--cycle;

\addplot[area legend, draw=black, fill=mylightgray3, forget plot]
table[row sep=crcr] {%
	x	y\\
	18.92	8.26\\
	21.26	8.27\\
	21.25	11.26\\
	18.96	11.2\\
}--cycle;

\addplot[area legend, draw=black, fill=mylightgray3, forget plot]
table[row sep=crcr] {%
	x	y\\
	25	7.7\\
	25.97	7.77\\
	26.15	11.26\\
	24.9	11.27\\
}--cycle;

\addlegendentry{$\mathrm{\small LoS}_1$}
\addlegendentry{$\mathrm{\small LoS}_2$}
\addlegendentry{$\mathrm{\small OLoS}_1$}
\addlegendentry{\!\!\!anchors}

\addplot [color=myblue, line width=\trajectoryLW, line join=round, line cap=round]
  table[row sep=crcr]{%
9.68073463439941	6.47091388702393\\
9.67666053771973	6.47262287139893\\
9.67242050170898	6.47597599029541\\
9.67074012756348	6.47742462158203\\
9.67230033874512	6.47675514221191\\
9.67433738708496	6.47631549835205\\
9.67476367950439	6.4775972366333\\
9.67445373535156	6.47875165939331\\
9.67445373535156	6.47874307632446\\
9.67375373840332	6.47820806503296\\
9.67338752746582	6.47718238830566\\
9.67344093322754	6.47749853134155\\
9.67100143432617	6.4790096282959\\
9.67095756530762	6.48020267486572\\
9.67087173461914	6.48216772079468\\
9.66915130615234	6.4818263053894\\
9.66953468322754	6.48168277740479\\
9.67044067382812	6.48218393325806\\
9.671630859375	6.48320484161377\\
9.67318344116211	6.48196029663086\\
9.67319869995117	6.48231220245361\\
9.67162132263184	6.4828577041626\\
9.67110252380371	6.48328590393066\\
9.67171669006348	6.48241329193115\\
9.67119026184082	6.4836049079895\\
9.66887092590332	6.48436641693115\\
9.67188930511475	6.48415374755859\\
9.67765998840332	6.48400831222534\\
9.68675136566162	6.4829740524292\\
9.70213317871094	6.47967529296875\\
9.71946144104004	6.47938632965088\\
9.7458324432373	6.47666263580322\\
9.77434158325195	6.47544860839844\\
9.80256080627441	6.4736909866333\\
9.83802890777588	6.47198343276978\\
9.8766040802002	6.46966075897217\\
9.9228515625	6.46828651428223\\
9.97006034851074	6.4656457901001\\
10.0178089141846	6.46438884735107\\
10.0685214996338	6.46323490142822\\
10.1216650009155	6.4636492729187\\
10.1749668121338	6.46190166473389\\
10.2340402603149	6.46086025238037\\
10.2879667282104	6.46183252334595\\
10.3428211212158	6.46125984191895\\
10.4026432037354	6.45817899703979\\
10.4644374847412	6.45537281036377\\
10.521635055542	6.4505500793457\\
10.5782108306885	6.44886779785156\\
10.6383056640625	6.44847345352173\\
10.7032318115234	6.4471607208252\\
10.7690162658691	6.44470548629761\\
10.8320274353027	6.44250011444092\\
10.897518157959	6.44389057159424\\
10.9642848968506	6.44656658172607\\
11.0318002700806	6.45086288452148\\
11.0983095169067	6.4566125869751\\
11.1659727096558	6.45649242401123\\
11.2316856384277	6.45645952224731\\
11.2973833084106	6.457444190979\\
11.3647565841675	6.45609188079834\\
11.4326763153076	6.45598602294922\\
11.5011672973633	6.46084833145142\\
11.5684947967529	6.46403026580811\\
11.6401195526123	6.47042274475098\\
11.7151126861572	6.47523593902588\\
11.789134979248	6.47597408294678\\
11.8600997924805	6.47680950164795\\
11.9283313751221	6.47818851470947\\
11.996129989624	6.47734022140503\\
12.0708618164062	6.47937917709351\\
12.1464519500732	6.48370361328125\\
12.2247905731201	6.48915576934814\\
12.3079929351807	6.49204206466675\\
12.3864097595215	6.49519205093384\\
12.4592952728271	6.49803066253662\\
12.5296115875244	6.50514221191406\\
12.5988540649414	6.5043511390686\\
12.673864364624	6.50555610656738\\
12.755859375	6.51167774200439\\
12.8478755950928	6.51770162582397\\
12.9247245788574	6.52436447143555\\
12.9996967315674	6.52901363372803\\
13.074520111084	6.52867603302002\\
13.1491527557373	6.52521991729736\\
13.2263412475586	6.52613401412964\\
13.3051071166992	6.52630138397217\\
13.3812618255615	6.52926206588745\\
13.4596633911133	6.53447151184082\\
13.5319957733154	6.53931140899658\\
13.6067771911621	6.54337501525879\\
13.6903324127197	6.5496621131897\\
13.7674217224121	6.55571937561035\\
13.8435230255127	6.56307029724121\\
13.9237442016602	6.56886100769043\\
14.004337310791	6.57853698730469\\
14.0795269012451	6.58883762359619\\
14.1585655212402	6.59821796417236\\
14.2344532012939	6.59961700439453\\
14.3100433349609	6.60484886169434\\
14.3831272125244	6.60931730270386\\
14.4588966369629	6.61321258544922\\
14.5369968414307	6.6203498840332\\
14.6187171936035	6.6313419342041\\
14.698600769043	6.64057350158691\\
14.7716445922852	6.64726257324219\\
14.8453483581543	6.65431308746338\\
14.9190521240234	6.65834617614746\\
14.9935703277588	6.66451358795166\\
15.0733585357666	6.66746234893799\\
15.1504974365234	6.67416000366211\\
15.2310371398926	6.67854118347168\\
15.3137512207031	6.68435287475586\\
15.3948497772217	6.68784189224243\\
15.4708499908447	6.68926811218262\\
15.5455894470215	6.69094181060791\\
15.6203155517578	6.69041061401367\\
15.6979866027832	6.68592357635498\\
15.7734546661377	6.68291854858398\\
15.849048614502	6.68387508392334\\
15.9210453033447	6.68462562561035\\
15.9910926818848	6.6860032081604\\
16.0611953735352	6.68754577636719\\
16.1348037719727	6.69136810302734\\
16.2125110626221	6.6953911781311\\
16.2907867431641	6.6977481842041\\
16.3702125549316	6.69463443756104\\
16.446475982666	6.69705963134766\\
16.5187206268311	6.69913053512573\\
16.5933017730713	6.70170116424561\\
16.6709747314453	6.69971370697021\\
16.74729347229	6.69648790359497\\
16.8253307342529	6.69527292251587\\
16.9030113220215	6.69964504241943\\
16.9786701202393	6.70084238052368\\
17.0537719726562	6.69822072982788\\
17.1258869171143	6.69397163391113\\
17.1974086761475	6.68941688537598\\
17.277494430542	6.68688440322876\\
17.3564033508301	6.68342208862305\\
17.4348068237305	6.67976856231689\\
17.5119934082031	6.67426776885986\\
17.5848217010498	6.66676092147827\\
17.6624641418457	6.65892553329468\\
17.7445468902588	6.65118789672852\\
17.8189239501953	6.6397533416748\\
17.8977470397949	6.63290309906006\\
17.9741535186768	6.62830018997192\\
18.046422958374	6.61926460266113\\
18.1211643218994	6.61398029327393\\
18.196382522583	6.61184883117676\\
18.2727546691895	6.60770130157471\\
18.3487529754639	6.60626220703125\\
18.4209213256836	6.6020245552063\\
18.497220993042	6.59690523147583\\
18.568977355957	6.58801794052124\\
18.6475486755371	6.58403205871582\\
18.7255687713623	6.58043479919434\\
18.8033638000488	6.57822418212891\\
18.8867454528809	6.57662487030029\\
18.9582920074463	6.57352256774902\\
19.027738571167	6.57007694244385\\
19.1003093719482	6.56306934356689\\
19.1728401184082	6.55829429626465\\
19.2495403289795	6.55682849884033\\
19.3234786987305	6.55306148529053\\
19.3951663970947	6.54971837997437\\
19.4709148406982	6.54561758041382\\
19.5430240631104	6.53964757919312\\
19.6247024536133	6.53496932983398\\
19.7033939361572	6.52903270721436\\
19.7814674377441	6.52893400192261\\
19.8557739257812	6.52998924255371\\
19.928373336792	6.53091430664062\\
19.9990749359131	6.53233814239502\\
20.0718154907227	6.52708625793457\\
20.1517505645752	6.52497339248657\\
20.2305679321289	6.51697158813477\\
20.3090324401855	6.51605939865112\\
20.3831596374512	6.51638507843018\\
20.4539070129395	6.51300525665283\\
20.5262279510498	6.50861072540283\\
20.6088104248047	6.50617599487305\\
20.6838893890381	6.50521516799927\\
20.7613849639893	6.50470447540283\\
20.8365917205811	6.50544595718384\\
20.9149227142334	6.50593137741089\\
20.9931907653809	6.50813293457031\\
21.0718593597412	6.50935840606689\\
21.1457347869873	6.51349830627441\\
21.2235412597656	6.51432418823242\\
21.3018779754639	6.51247072219849\\
21.3796539306641	6.51157188415527\\
21.4563331604004	6.5120267868042\\
21.5317726135254	6.51568269729614\\
21.6045379638672	6.52130794525146\\
21.6776676177979	6.52387475967407\\
21.7576713562012	6.52771472930908\\
21.8367328643799	6.53096580505371\\
21.9155502319336	6.53292417526245\\
21.9936294555664	6.53513336181641\\
22.0720672607422	6.54016733169556\\
22.1535701751709	6.54211139678955\\
22.230562210083	6.54338216781616\\
22.3078689575195	6.54581546783447\\
22.3840274810791	6.54807376861572\\
22.4600162506104	6.55200672149658\\
22.5355091094971	6.55427742004395\\
22.6140193939209	6.55545520782471\\
22.6941719055176	6.55921077728271\\
22.7705307006836	6.56279182434082\\
22.849588394165	6.56528663635254\\
22.9238014221191	6.56499099731445\\
22.9956150054932	6.56959819793701\\
23.0694923400879	6.5761022567749\\
23.1424198150635	6.57801198959351\\
23.2194881439209	6.5800142288208\\
23.2998237609863	6.5797872543335\\
23.3773632049561	6.58057737350464\\
23.4507484436035	6.58398818969727\\
23.5209217071533	6.58662748336792\\
23.5968475341797	6.58936786651611\\
23.6681060791016	6.5942268371582\\
23.7445087432861	6.59657192230225\\
23.8220500946045	6.59940814971924\\
23.8983383178711	6.6053729057312\\
23.9760112762451	6.60818290710449\\
24.0515613555908	6.61086177825928\\
24.1284427642822	6.6103982925415\\
24.2028884887695	6.61379051208496\\
24.2774429321289	6.61382722854614\\
24.3526668548584	6.61519479751587\\
24.4248199462891	6.61704730987549\\
24.4986534118652	6.61885499954224\\
24.5750350952148	6.62178945541382\\
24.6580696105957	6.62624025344849\\
24.7371292114258	6.63049745559692\\
24.820369720459	6.6355504989624\\
24.8979244232178	6.6396312713623\\
24.9717464447021	6.6447057723999\\
25.0419292449951	6.64359569549561\\
25.1102809906006	6.63966608047485\\
25.1839828491211	6.64441919326782\\
25.2602024078369	6.63621997833252\\
25.3416881561279	6.62472534179688\\
25.4011611938477	6.61729574203491\\
25.469747543335	6.60689449310303\\
25.5367469787598	6.60415363311768\\
25.6088562011719	6.59681415557861\\
25.6828174591064	6.59271335601807\\
25.7531604766846	6.58936738967896\\
25.8183784484863	6.58190298080444\\
25.8841667175293	6.56732559204102\\
25.9552841186523	6.55012226104736\\
26.0254573822021	6.53263330459595\\
26.0954494476318	6.51410102844238\\
26.1649188995361	6.49483871459961\\
26.2353286743164	6.47240734100342\\
26.3051414489746	6.451340675354\\
26.3812160491943	6.42742347717285\\
26.4482517242432	6.4026894569397\\
26.5169372558594	6.3732852935791\\
26.5845737457275	6.34111499786377\\
26.6528339385986	6.31159830093384\\
26.71728515625	6.28785085678101\\
26.7790546417236	6.26325130462646\\
26.8392295837402	6.24582004547119\\
26.8914909362793	6.23052358627319\\
26.9430561065674	6.21900367736816\\
26.9908351898193	6.20962905883789\\
27.0349884033203	6.20264911651611\\
27.0783863067627	6.20006990432739\\
27.119161605835	6.20161771774292\\
27.1571292877197	6.20562696456909\\
27.1911468505859	6.20987606048584\\
27.2232456207275	6.21590185165405\\
27.2524509429932	6.22600984573364\\
27.2842350006104	6.23397970199585\\
27.3159427642822	6.24755954742432\\
27.3482933044434	6.26161766052246\\
27.3782176971436	6.27593469619751\\
27.4064693450928	6.29008531570435\\
27.4331302642822	6.30362272262573\\
27.4582271575928	6.31924724578857\\
27.4807739257812	6.33469009399414\\
27.499174118042	6.35129880905151\\
27.5173492431641	6.36697292327881\\
27.5244979858398	6.38174438476562\\
27.5272159576416	6.39270210266113\\
27.5286731719971	6.40703964233398\\
27.5240287780762	6.41767597198486\\
27.5221462249756	6.4225959777832\\
27.5203056335449	6.43009376525879\\
27.5193157196045	6.43743515014648\\
27.5200748443604	6.44670867919922\\
27.523042678833	6.46243524551392\\
27.5217380523682	6.47716236114502\\
27.5171642303467	6.49631404876709\\
27.51438331604	6.51452684402466\\
27.512336730957	6.53247928619385\\
27.5078716278076	6.55544376373291\\
27.5029220581055	6.57113933563232\\
27.4970664978027	6.58502960205078\\
27.4897747039795	6.59951114654541\\
27.4845294952393	6.60850048065186\\
27.4769954681396	6.61577701568604\\
27.4671001434326	6.62562417984009\\
27.4608345031738	6.6342945098877\\
27.451696395874	6.64532327651978\\
27.4448108673096	6.6589503288269\\
27.4369964599609	6.67202472686768\\
27.4280529022217	6.68498039245605\\
27.4126300811768	6.70092582702637\\
27.3970165252686	6.71446800231934\\
27.3839645385742	6.72955322265625\\
27.3644199371338	6.74073076248169\\
27.3411617279053	6.7536096572876\\
27.3171844482422	6.76587677001953\\
27.2998600006104	6.77915191650391\\
27.277322769165	6.79432678222656\\
27.2566795349121	6.80411148071289\\
27.236141204834	6.81569671630859\\
27.2123203277588	6.82450103759766\\
27.1903858184814	6.83097076416016\\
27.1659488677979	6.8361349105835\\
27.1396617889404	6.83364915847778\\
27.115592956543	6.83453845977783\\
27.0939331054688	6.84021377563477\\
27.0681209564209	6.83960437774658\\
27.0406646728516	6.83911466598511\\
27.01047706604	6.84106492996216\\
26.976318359375	6.83678817749023\\
26.9411201477051	6.82917308807373\\
26.904598236084	6.82301807403564\\
26.8667221069336	6.81073474884033\\
26.8210945129395	6.79848146438599\\
26.7697486877441	6.7831768989563\\
26.7214698791504	6.76761436462402\\
26.6718196868896	6.75309562683105\\
26.6205863952637	6.73869180679321\\
26.5711345672607	6.72148132324219\\
26.5156173706055	6.70275688171387\\
26.456226348877	6.68207550048828\\
26.3999691009521	6.65684795379639\\
26.3415184020996	6.63440608978271\\
26.2861518859863	6.60620975494385\\
26.2271575927734	6.57747840881348\\
26.1660041809082	6.54807281494141\\
26.1114177703857	6.52201175689697\\
26.0532112121582	6.4886326789856\\
25.9942665100098	6.45505905151367\\
25.9287185668945	6.42185401916504\\
25.8632621765137	6.38432884216309\\
25.793083190918	6.35159015655518\\
25.7167053222656	6.32160425186157\\
25.647346496582	6.29796266555786\\
25.5827560424805	6.27388334274292\\
25.5232563018799	6.2506160736084\\
25.4658527374268	6.22086477279663\\
25.4087314605713	6.19895362854004\\
25.3574886322021	6.179762840271\\
25.3151416778564	6.16015672683716\\
25.2621574401855	6.13874435424805\\
25.2154083251953	6.11777544021606\\
25.1666240692139	6.09752082824707\\
25.1175765991211	6.08019828796387\\
25.0669631958008	6.06288051605225\\
25.026050567627	6.04641056060791\\
24.9828128814697	6.03144884109497\\
24.9328441619873	6.0126781463623\\
24.8780918121338	6.00133895874023\\
24.8226680755615	5.98889589309692\\
24.7635974884033	5.97793960571289\\
24.700325012207	5.96876573562622\\
24.6310787200928	5.95663118362427\\
24.5591049194336	5.94502019882202\\
24.4879531860352	5.93549728393555\\
24.41845703125	5.92609739303589\\
24.3503189086914	5.92451953887939\\
24.286828994751	5.92134714126587\\
24.2142868041992	5.9189624786377\\
24.1415100097656	5.91839075088501\\
24.0692443847656	5.91378116607666\\
23.9916286468506	5.91227912902832\\
23.9105987548828	5.91075420379639\\
23.8371829986572	5.91213035583496\\
23.7636528015137	5.91018199920654\\
23.6922187805176	5.91189765930176\\
23.6221294403076	5.91325569152832\\
23.5520076751709	5.91404867172241\\
23.4803218841553	5.91565370559692\\
23.4081535339355	5.91608142852783\\
23.3305816650391	5.91809892654419\\
23.2493877410889	5.91976594924927\\
23.1667823791504	5.92133808135986\\
23.0896320343018	5.91833257675171\\
23.011194229126	5.91670799255371\\
22.9366912841797	5.910804271698\\
22.8628616333008	5.91137218475342\\
22.7857513427734	5.91142225265503\\
22.7135810852051	5.91077756881714\\
22.6410961151123	5.9104471206665\\
22.5600833892822	5.90958213806152\\
22.4788398742676	5.90608358383179\\
22.4079322814941	5.90178108215332\\
22.3336868286133	5.89909076690674\\
22.2594509124756	5.90041399002075\\
22.1810626983643	5.90208148956299\\
22.1002750396729	5.90079355239868\\
22.0221366882324	5.89938306808472\\
21.9459247589111	5.89685916900635\\
21.8745670318604	5.89504671096802\\
21.7981624603271	5.89507436752319\\
21.7197093963623	5.89687204360962\\
21.643533706665	5.90220499038696\\
21.5698699951172	5.90632963180542\\
21.4906768798828	5.91053199768066\\
21.4150257110596	5.91517353057861\\
21.3386554718018	5.92045259475708\\
21.2667293548584	5.92531251907349\\
21.1880874633789	5.9248366355896\\
21.1076221466064	5.92545223236084\\
21.0309734344482	5.92512941360474\\
20.9549694061279	5.930495262146\\
20.8785285949707	5.93244981765747\\
20.7993946075439	5.93961381912231\\
20.730354309082	5.94463157653809\\
20.6619052886963	5.94598197937012\\
20.5852355957031	5.94966268539429\\
20.5084457397461	5.95352649688721\\
20.4392223358154	5.95783090591431\\
20.3625335693359	5.96704435348511\\
20.2915878295898	5.97153520584106\\
20.2220211029053	5.97587108612061\\
20.1498622894287	5.9855694770813\\
20.0730171203613	5.99505472183228\\
20.0014457702637	6.00073575973511\\
19.9291305541992	6.00591516494751\\
19.8572731018066	6.01213645935059\\
19.7808780670166	6.01740312576294\\
19.6975212097168	6.02486944198608\\
19.6167602539062	6.03665924072266\\
19.5315780639648	6.04655694961548\\
19.4477481842041	6.0553560256958\\
19.3637771606445	6.06883859634399\\
19.2861175537109	6.08069038391113\\
19.2094955444336	6.08787631988525\\
19.1313095092773	6.09625244140625\\
19.0560874938965	6.10528564453125\\
18.9818382263184	6.11294555664062\\
18.9117813110352	6.12160158157349\\
18.8378314971924	6.13068294525146\\
18.7700500488281	6.1406102180481\\
18.6909198760986	6.15094566345215\\
18.6157093048096	6.15817499160767\\
18.5401344299316	6.16769790649414\\
18.4658660888672	6.1784815788269\\
18.3916606903076	6.1869044303894\\
18.3152904510498	6.1977596282959\\
18.2357044219971	6.20677661895752\\
18.1610012054443	6.21747303009033\\
18.0908203125	6.22946834564209\\
18.0209083557129	6.24117469787598\\
17.9494171142578	6.25685453414917\\
17.8788032531738	6.27079677581787\\
17.8083457946777	6.28512239456177\\
17.735034942627	6.29720592498779\\
17.6561431884766	6.30964040756226\\
17.5844116210938	6.32231187820435\\
17.5087108612061	6.33346462249756\\
17.4330348968506	6.34668445587158\\
17.3636989593506	6.35742473602295\\
17.2890491485596	6.36711692810059\\
17.2152576446533	6.37457847595215\\
17.1380577087402	6.37894058227539\\
17.0694961547852	6.38186168670654\\
17.0015201568604	6.38633251190186\\
16.9295692443848	6.39029121398926\\
16.8616371154785	6.39269399642944\\
16.7856464385986	6.38883590698242\\
16.7122955322266	6.38545608520508\\
16.6400737762451	6.38507270812988\\
16.5650520324707	6.38343667984009\\
16.4947624206543	6.38006353378296\\
16.4209823608398	6.3794093132019\\
16.3421192169189	6.37158250808716\\
16.2710037231445	6.36483764648438\\
16.1970443725586	6.35548686981201\\
16.1217918395996	6.34866046905518\\
16.0520973205566	6.33759498596191\\
15.9785842895508	6.32463073730469\\
15.9027156829834	6.31360912322998\\
15.8307628631592	6.30152893066406\\
15.7562713623047	6.29186677932739\\
15.6898899078369	6.28099536895752\\
15.628625869751	6.26519775390625\\
15.557258605957	6.2497181892395\\
15.486400604248	6.23489999771118\\
15.4115905761719	6.21841859817505\\
15.3394680023193	6.2009449005127\\
15.2696018218994	6.18700122833252\\
15.1967868804932	6.17035007476807\\
15.1211624145508	6.15263843536377\\
15.047966003418	6.13057470321655\\
14.9761142730713	6.11094379425049\\
14.9038105010986	6.0967845916748\\
14.8283615112305	6.08125972747803\\
14.7586860656738	6.06315946578979\\
14.6844882965088	6.04672718048096\\
14.6065864562988	6.02896308898926\\
14.5326862335205	6.01380586624146\\
14.4550228118896	5.99366664886475\\
14.3767871856689	5.97935056686401\\
14.3001232147217	5.96716499328613\\
14.2255115509033	5.95377254486084\\
14.1501846313477	5.9398889541626\\
14.0729675292969	5.92664480209351\\
13.9982509613037	5.91501760482788\\
13.9186992645264	5.90239763259888\\
13.8398532867432	5.89236211776733\\
13.758752822876	5.88082313537598\\
13.6767616271973	5.86986446380615\\
13.5996551513672	5.86296224594116\\
13.5219974517822	5.85365390777588\\
13.4412078857422	5.84737873077393\\
13.3589649200439	5.84680080413818\\
13.2719974517822	5.84626817703247\\
13.1903228759766	5.84498071670532\\
13.1165752410889	5.85146045684814\\
13.0448570251465	5.85480642318726\\
12.9656620025635	5.85917901992798\\
12.8965721130371	5.86003828048706\\
12.822416305542	5.86113357543945\\
12.7447967529297	5.8652081489563\\
12.677713394165	5.87280893325806\\
12.6009712219238	5.87795400619507\\
12.5275096893311	5.88520097732544\\
12.4518928527832	5.88942575454712\\
12.3768501281738	5.90059995651245\\
12.3045883178711	5.90992021560669\\
12.234375	5.91914653778076\\
12.1650371551514	5.92863893508911\\
12.0972480773926	5.9445538520813\\
12.025218963623	5.95369434356689\\
11.9458465576172	5.96701335906982\\
11.8746871948242	5.97795391082764\\
11.802131652832	5.9911003112793\\
11.7269992828369	6.00283527374268\\
11.6447505950928	6.01352787017822\\
11.5601387023926	6.02680540084839\\
11.4807147979736	6.03928804397583\\
11.4061460494995	6.05521631240845\\
11.3276882171631	6.06759977340698\\
11.2536010742188	6.07929420471191\\
11.1807136535645	6.091468334198\\
11.1073818206787	6.1003737449646\\
11.034984588623	6.11378049850464\\
10.9594593048096	6.12141084671021\\
10.889835357666	6.1271858215332\\
10.8178281784058	6.13001108169556\\
10.7445764541626	6.13271856307983\\
10.6719722747803	6.13323783874512\\
10.5973720550537	6.13206100463867\\
10.5228958129883	6.1351203918457\\
10.4540281295776	6.13700771331787\\
10.3877334594727	6.13219499588013\\
10.3298568725586	6.12699222564697\\
10.278636932373	6.12431859970093\\
10.2314567565918	6.12258768081665\\
10.1906747817993	6.11895656585693\\
10.1540241241455	6.11427545547485\\
10.1210260391235	6.10940599441528\\
10.0825538635254	6.10429191589355\\
10.0439281463623	6.10000467300415\\
10.0062465667725	6.10176372528076\\
9.97172164916992	6.10130214691162\\
9.93888664245605	6.09835338592529\\
9.90441226959229	6.09435939788818\\
9.86771774291992	6.09127283096313\\
9.83525657653809	6.08696365356445\\
9.80827331542969	6.08465242385864\\
9.7840576171875	6.07998037338257\\
9.76296997070312	6.08010101318359\\
9.74269485473633	6.08003425598145\\
9.7328634262085	6.08010149002075\\
9.72714614868164	6.07950735092163\\
9.72225761413574	6.08033323287964\\
9.7216911315918	6.07997465133667\\
9.71999168395996	6.08071517944336\\
9.72070980072021	6.08147525787354\\
9.72340488433838	6.08194637298584\\
9.71897506713867	6.08145618438721\\
9.71821784973145	6.08054447174072\\
9.71843147277832	6.08294057846069\\
9.71949863433838	6.0814061164856\\
9.71979331970215	6.080810546875\\
9.72200489044189	6.07939386367798\\
9.7208309173584	6.0788950920105\\
9.72147178649902	6.07961463928223\\
9.7213020324707	6.08023452758789\\
9.7243766784668	6.08164405822754\\
9.72361850738525	6.08155250549316\\
9.72222518920898	6.08176183700562\\
9.71978855133057	6.08070993423462\\
9.72194480895996	6.08115816116333\\
9.72051334381104	6.08015966415405\\
9.72208595275879	6.0789270401001\\
9.72255325317383	6.07881641387939\\
};\label{pgf:Trackref}

\addplot [color=mylightblue3, dashed, line width=\trajectoryLW, line join=round, line cap=round]
  table[row sep=crcr]{%
9.62800623323308	6.48856771837916\\
9.63852970765504	6.4915016633861\\
9.93094528848375	6.48182069779417\\
10.064592728734	6.47603618011515\\
10.2667680411947	6.46700143951794\\
10.4478494995981	6.46001337549104\\
10.5286933420989	6.46065232475761\\
10.7068530285771	6.46022005709768\\
10.8289993221635	6.45927777237429\\
11.0214292814132	6.46449806159781\\
11.1290392264618	6.46565968203918\\
11.4033472788757	6.46785574661603\\
11.5341625438844	6.47251420819241\\
11.7561127661531	6.48237877871477\\
11.9753460036554	6.48271780739411\\
12.1958312177592	6.49130989552842\\
12.327890163832	6.50099068592712\\
12.4233929228805	6.50490397824012\\
12.5737469038838	6.51884508977691\\
12.7862611837568	6.52360322653028\\
13.1138800978093	6.5328494992261\\
13.2475029316631	6.53325290411834\\
13.4297323807376	6.54288276941383\\
13.5245797366951	6.55053084783832\\
13.6379179527055	6.55829960241218\\
13.8469365250436	6.56983064069746\\
14.0729239333848	6.59101553548913\\
14.1918853090465	6.60196664217955\\
14.5408202677146	6.62735511487705\\
14.7090078234696	6.64001989188108\\
15.0299953155267	6.65285899695179\\
15.266487083508	6.65878960164268\\
15.3514902619777	6.65971393681293\\
15.5149362992816	6.65241235663408\\
15.613668362881	6.65010045786925\\
15.7021304156246	6.64584536507118\\
15.9416436018324	6.62844294779342\\
16.1585208547376	6.62426616525019\\
16.2838187114762	6.62055747465276\\
16.451675866985	6.6158413788683\\
16.5843428953821	6.60746065170146\\
16.9394110121182	6.58111767782238\\
17.0338786752658	6.57775508599428\\
17.2323273349929	6.56777463279478\\
17.5241437627603	6.5571498441657\\
17.6264173736262	6.55396179404054\\
17.840617600478	6.55058972869776\\
18.0045704898524	6.54448695564738\\
18.1457258463582	6.54054288963269\\
18.4259524889478	6.5387278598732\\
18.7938626297506	6.54717072728941\\
18.8923701387896	6.55018358342381\\
19.0812726060454	6.54795948439416\\
19.5500920768156	6.53003015563673\\
19.979066706736	6.52339844387397\\
20.0799946341286	6.52091674304251\\
20.2531748034584	6.51534018225853\\
20.4550876831934	6.51666121831513\\
20.5904692961796	6.51536400154662\\
20.7395831935336	6.51690004214079\\
20.8459228466208	6.5189953173983\\
21.0587371769719	6.51705577779025\\
21.3554248965919	6.51842110887612\\
21.6232808497159	6.51482696861862\\
21.7480003126964	6.51671174555193\\
21.9159339289751	6.52574416136436\\
22.0218353284508	6.52847232805651\\
22.1317793555948	6.5322685374438\\
22.2678950575465	6.53519382032318\\
22.8743994478515	6.54225107451478\\
23.1309521173729	6.54766831412288\\
23.2497067870572	6.54532274852957\\
23.3412855449564	6.54393455726878\\
23.9483392520255	6.54584815425686\\
24.1737501236769	6.53637710770994\\
24.2603573042865	6.53194762617551\\
24.6773831461049	6.51062942485646\\
25.1247082003947	6.46497152564308\\
25.2393714121195	6.44385351754526\\
25.3554917938222	6.42370810135417\\
25.5513390853176	6.3969490451476\\
25.6688005505583	6.38033275624101\\
25.7343504043464	6.37381497269273\\
25.8060697767183	6.36528055239198\\
25.9318773652132	6.34860620656889\\
26.1162192837539	6.32109289420091\\
26.219575858187	6.29919396105829\\
26.7192297468975	6.18636005568359\\
27.0983162864818	6.12692323648666\\
27.151122166478	6.12753057559226\\
27.2358263724092	6.14069486976693\\
27.3929996144926	6.18320468096511\\
27.4355294083562	6.20022199177938\\
27.5455581413686	6.25729035410929\\
27.6053896076074	6.29757493464171\\
27.6369326063683	6.34318962480587\\
27.6622587485309	6.39524043246098\\
27.679799020054	6.42546175259995\\
27.7238245577514	6.53318502430773\\
27.7338664512668	6.55897244662905\\
27.7430227980356	6.60489532818742\\
27.7557140037978	6.68458776223353\\
27.750303912927	6.71962261133936\\
27.7414521250632	6.75258860348422\\
27.7389093265283	6.77426434125596\\
27.7243978145448	6.80925516075495\\
27.7013998812512	6.84242705176255\\
27.680726613137	6.87034675399459\\
27.65477709829	6.90520175961074\\
27.6282012706648	6.94758035655364\\
27.60557261884	6.96714977853989\\
27.5053427589786	7.03885523031319\\
27.4703369503874	7.05628586868762\\
27.3894792069281	7.0825233794985\\
27.3494886456122	7.09018620868013\\
27.2211654423981	7.08976306048296\\
27.1360958318055	7.08563159099547\\
27.0251390229255	7.07808284304216\\
26.9672716979238	7.06888828684496\\
26.7635421049951	7.02569792871988\\
26.4365384529823	6.93777305244306\\
26.3576920965607	6.91361442784433\\
25.8919267677002	6.77354612535524\\
25.7814208701418	6.74296515817241\\
25.6878783214663	6.71092592618803\\
25.5826804287396	6.67077475192414\\
25.442688819764	6.62507156700486\\
25.3552781197545	6.59615724055854\\
24.9709961836945	6.47287255292854\\
24.7733645150749	6.40119425824083\\
24.5435409475929	6.32182253568479\\
24.4799318415288	6.29593090221435\\
24.3385255509851	6.23492410883229\\
24.2404386507846	6.20065172534897\\
24.1064976335759	6.15749713019299\\
24.0211476223944	6.13554243111368\\
23.878601841345	6.10512567552303\\
23.6348817347481	6.0617880651143\\
23.5192672720284	6.04989628427732\\
23.4285171132881	6.03725798554051\\
23.3532659090903	6.02995681312459\\
23.2613711419578	6.02717038146768\\
23.0457911935965	6.02820539163665\\
22.8600139434313	6.02991534935073\\
22.4696698933271	6.01093227150601\\
22.3807324440502	6.00744247714119\\
22.0736497567087	5.99972119763303\\
21.5696596653162	5.98707620223662\\
21.4549417499603	5.98788420053334\\
21.2297855827076	5.98046259092053\\
21.1470731154652	5.97453889249176\\
20.5500106233526	5.98025947085502\\
20.4390006069291	5.98183346386678\\
20.274800947992	5.99429590624235\\
19.8368081141895	6.02666819389239\\
19.7455667837345	6.03697776653699\\
19.5811266721369	6.05515575124861\\
19.4268860765989	6.07086250181054\\
19.3567879396479	6.07974243917877\\
19.2855843395007	6.08793269476928\\
19.2053862768322	6.09637316272592\\
18.9841599523159	6.12764389255743\\
18.8756949073418	6.1409176401394\\
18.4025722753079	6.19403800402307\\
18.2512754718293	6.2117989111272\\
18.0901219881033	6.23150022800835\\
17.8358448922097	6.26263646013915\\
17.6705436874061	6.27361219444808\\
17.4993812879414	6.28442666971374\\
17.3759988993615	6.29472465384066\\
17.2735143198794	6.29853789805064\\
17.094049668676	6.29769465702367\\
17.0015801358569	6.29299877736856\\
16.8151745086331	6.27862284741623\\
16.4809519480301	6.2611736986875\\
16.3227802928541	6.24210710917472\\
16.2192526440648	6.2209345194067\\
16.1103482215898	6.19889158569213\\
15.9000311409556	6.16724746812629\\
15.7132343985953	6.13972099144886\\
15.4238747905943	6.0837495282016\\
15.1975227012868	6.03895125195323\\
15.0780307114863	6.01772998881204\\
14.9533471326839	5.99626994843348\\
14.8192052142496	5.9745599046318\\
14.7098324053177	5.95380692618692\\
14.4651330174044	5.92225431197214\\
14.2558657501729	5.89541109628538\\
14.1499179465637	5.88567780317763\\
13.9702383472586	5.87043355517405\\
13.8459318689862	5.85623145256962\\
13.7180174661225	5.84268302014472\\
13.5671208909544	5.83341404961816\\
13.4228459705832	5.83090637295285\\
13.0335278259046	5.82990814489443\\
12.9063607688167	5.83388149559833\\
12.4074089162803	5.85339852416593\\
12.085165465928	5.88299112224028\\
11.5255997962832	5.97348830019718\\
11.3830041356121	6.0018709644876\\
11.1806685923934	6.05260207363403\\
11.0552592201858	6.08702645648879\\
10.8649108706205	6.12753955181644\\
10.7674455361289	6.14342214941925\\
10.6839819588191	6.15112186848957\\
10.4887221792022	6.15839474688206\\
10.2978243190543	6.14850698839579\\
10.052825806787	6.1246976929915\\
9.9848953678592	6.11688592164629\\
9.82904227550201	6.10417886603144\\
9.73667642974931	6.10033266000857\\
9.68221003871126	6.10218005487493\\
9.6320476173549	6.10627834901998\\
9.64162024371921	6.1007056175463\\
9.63767168133577	6.1021593972112\\
};\label{pgf:Trackref3}

\addplot [color=myaltblue, line width=\trajectoryLW, line join=round, line cap=round]
  table[row sep=crcr]{%
9.7053758140998	6.51621495424516\\
9.71966631644507	6.51666176317031\\
9.84373670016276	6.51580438965158\\
10.164058945584	6.51054555033906\\
10.4401169498169	6.5111841902788\\
10.6665852225809	6.5284558280874\\
10.7782295629793	6.5335830564072\\
10.8749631679426	6.54086219881263\\
11.0778203327525	6.56791699770726\\
11.4421195040447	6.61951542326704\\
11.9020419620209	6.70331854894336\\
12.080631970176	6.73524784002092\\
12.2738523562515	6.77212647675678\\
12.5071979192291	6.82850560492022\\
12.658555037759	6.86010240750624\\
12.7753369022277	6.88482450256341\\
12.8613713113505	6.91045384721571\\
12.967590512036	6.9400677529311\\
13.1372854422351	6.9933183608227\\
13.2991628086648	7.05980655824932\\
13.3876321767487	7.10176717876382\\
13.5037351143117	7.15293229559305\\
13.5757883553949	7.18511735526226\\
13.7169663874426	7.26982997639568\\
13.9330191423974	7.42347929442275\\
13.9729085747716	7.46322832979395\\
14.080610753105	7.60003753511898\\
14.1256306325641	7.6633705078283\\
14.1794723101082	7.75999735048401\\
14.2414809389898	7.88664397380096\\
14.269651032035	7.95644508249518\\
14.3576593089109	8.23634312002248\\
14.3747766879687	8.31561402401558\\
14.3897767134835	8.4250823081967\\
14.4055373252888	8.79551587500738\\
14.4107624304267	8.9001928282619\\
14.4121717940765	9.05403582610312\\
14.4095089834026	9.15961731541596\\
14.3999380072332	9.29495403332329\\
14.3941588683132	9.41410697012927\\
14.3829179638426	9.57893509941308\\
14.3837060416486	9.67620481995232\\
14.3749061274169	10.0080313318177\\
14.3793135616161	10.2041643954981\\
14.3753607778996	10.2659876256975\\
14.3684899394207	10.3508652885279\\
14.3569481263343	10.7524949237194\\
14.3650168870512	10.9928728868423\\
14.3679803209908	11.0665640637014\\
14.3636250865662	11.4019157335217\\
14.3647593368543	11.6149860339536\\
14.3568272319364	11.7469960876301\\
14.3556844364904	11.8554857562868\\
14.3434281681396	11.9434046927621\\
14.3334228705052	12.0212393598598\\
14.3298618181922	12.3069153644373\\
14.3351969675066	12.5299061259393\\
14.3308465448256	12.5877564371607\\
14.3349882588798	12.720136560535\\
14.327149689054	12.8137255212911\\
14.3048787418644	12.9942899373048\\
14.2799544325074	13.1551323535727\\
14.2588424152613	13.2738430194334\\
14.2444429997562	13.366176646994\\
14.2357366095561	13.4232984306292\\
14.2268113783718	13.4913409938395\\
14.2004727913248	13.6087614099096\\
14.1853975888765	13.6636453208074\\
14.147353758297	13.7471450449333\\
14.1091241891452	13.8123570455051\\
14.0908307721084	13.8303719421215\\
14.0313715974337	13.8824524930911\\
14.004399125666	13.9046019442403\\
13.9501005185929	13.9424577444947\\
13.9021080635743	13.9770668607884\\
13.86290607275	13.9988580995041\\
13.7798701139599	14.0301050973401\\
13.7248163562535	14.0505428226187\\
13.6958962529824	14.0591125820693\\
13.5696914355395	14.0766530950872\\
13.5300639722289	14.0839717343592\\
13.4461953689857	14.096545346232\\
13.3470503530298	14.1094829024558\\
13.1958136763517	14.1211628296246\\
13.1191039617996	14.1202087947378\\
13.0480802735105	14.1185199561907\\
12.9362578973873	14.1146487356637\\
12.8139586257419	14.1117617326533\\
12.5059746486761	14.1100485717757\\
12.3549483515747	14.1032824770864\\
12.11555118969	14.1166890471553\\
11.9959409154847	14.1293135839744\\
11.6929432482862	14.165651151824\\
11.3541679141043	14.1987314053751\\
11.2125416447222	14.2172498462737\\
11.1066799825364	14.2232803966493\\
10.8215913243504	14.259578390248\\
10.4675237369053	14.3168078245909\\
10.3534620185603	14.327819834919\\
10.2867944397281	14.3262604758418\\
10.2155056815291	14.3174832746755\\
10.144471870874	14.3086886829902\\
10.0538631372082	14.3014235721488\\
9.97903439651239	14.2942577001773\\
9.88301944476997	14.2524907525551\\
9.81736501861623	14.2100289936271\\
9.79304766767042	14.1888637927328\\
9.74637195414426	14.1585657200632\\
9.70748537565739	14.1166551405131\\
9.63523601388012	14.047249066681\\
9.60511763412132	14.0129832079753\\
9.50104381457972	13.8748072591581\\
9.46517876062967	13.8074961630398\\
9.41663382707613	13.7101861508258\\
9.38852573881163	13.6467537817313\\
9.34902105057868	13.5259177666296\\
9.34161019019681	13.4749900727563\\
9.32903516977376	13.4015782354634\\
9.2954382148431	13.2058504076367\\
9.28306741062137	12.9890346399522\\
9.27798767223072	12.9034026645692\\
9.28051903854004	12.8065972149835\\
9.28990651521496	12.7252087689789\\
9.31195837526521	12.5669262971702\\
9.32107559085295	12.3761115084347\\
9.31954352378818	12.2817161029228\\
9.32508578093007	12.2207806620868\\
9.33789612661776	12.0967472447229\\
9.34078259367861	12.018873907727\\
9.34269696164754	11.8154087637698\\
9.34893687213064	11.6655474100642\\
9.34401898850381	11.5765374868066\\
9.30019830187712	11.1100169152169\\
9.28283697683074	11.0089814224767\\
9.20495785840827	10.6100951334923\\
9.19173613041382	10.5133980351758\\
9.17860961863626	10.3332897293302\\
9.18399975734152	10.2275143557508\\
9.19307695662533	10.1187620183597\\
9.20236938240387	10.0427010599232\\
9.23855980100825	9.81838505953182\\
9.25892911533395	9.70088103115266\\
9.27335313471066	9.6088798717287\\
9.33893836412817	9.34973073581351\\
9.36471891372523	9.27602038844199\\
9.57711113822646	8.75844448591701\\
9.63078605058486	8.6422215095515\\
9.68619665830278	8.52154300957562\\
9.79447406464381	8.29577123432058\\
9.86172835035646	8.16807831719767\\
9.97424091033272	7.95061925520981\\
10.0659022753564	7.78118893611686\\
10.2648268889782	7.41859971236102\\
10.3634292360384	7.26734669977654\\
10.5278052681949	7.0360647962758\\
10.8273151370034	6.70118583756681\\
10.907889996207	6.63333112560709\\
11.0537874511268	6.52181874674984\\
11.1212192950984	6.48057499593141\\
11.2310199721312	6.40880327074621\\
11.2679051891197	6.38833317050677\\
11.3353623701668	6.36482371720366\\
11.3950031140214	6.34946215825131\\
11.5886327380277	6.31664140629499\\
11.6744374714742	6.30442332465354\\
11.7793798452862	6.29392182490957\\
11.9530293981362	6.28295773910599\\
12.0727730384286	6.28369734992557\\
12.1468783299879	6.29368290371177\\
12.2351803880733	6.30403228542205\\
12.4604718746652	6.32520051402591\\
12.7354583037853	6.35893843483873\\
12.8695423300649	6.38967242140105\\
12.9607275925509	6.41324074240367\\
13.0553495706778	6.44120520746376\\
13.1586865004364	6.48248963538614\\
13.2341913655395	6.52118175260254\\
13.3826396806663	6.60824070545832\\
13.4520667329455	6.66117687265089\\
13.5579397865766	6.75246126992139\\
13.64281033522	6.83041786979346\\
13.7142544244983	6.90911194533584\\
13.7653190646384	6.97479532877385\\
13.8079246225856	7.03057392328851\\
13.8553501487051	7.10876775760444\\
13.9109477014104	7.2066154136132\\
13.9396155385499	7.27347708955236\\
13.9610629147138	7.34406118448719\\
14.0000645607258	7.4899922381097\\
14.0263566665733	7.78750142127517\\
14.0309928919097	7.91904231820982\\
14.0273272569447	8.18507312109905\\
14.0178647602418	8.69300083797548\\
13.9947921587182	9.07861422630747\\
13.9851000039511	9.19578052308785\\
13.9745820778727	9.38400442916557\\
13.9570965617646	9.51960186593421\\
13.9424826443282	9.60142306768847\\
13.9206819534792	9.71523657658556\\
13.9030020605353	9.78399975751317\\
13.8702721010272	9.91192989816013\\
13.8458522317566	10.0299484371368\\
13.8395240153389	10.1039541030962\\
13.8358711341938	10.2412293612134\\
13.8345093363624	10.8548072044504\\
13.8521056819768	11.1100749708621\\
13.8681933208299	11.2035185384388\\
13.9249289476128	11.4556157995375\\
14.0549437977626	11.8781321152102\\
14.0747463545309	11.9468135871507\\
14.0934903052028	12.0533054083677\\
14.1795959004952	12.5190340109426\\
14.2033173758983	12.6157333446144\\
14.2095685772154	12.6558266982043\\
14.2435290826533	12.8139489317085\\
14.244572694942	12.8738908293105\\
14.2323703563883	13.2000644998371\\
14.2228363314531	13.3319927414631\\
14.1999759065197	13.4776587986346\\
14.1722711071548	13.5887741016388\\
14.1229330471673	13.7055977152587\\
14.0645282674979	13.8061775232612\\
13.9788037095192	13.913595429852\\
13.9227005540118	13.9637788176222\\
13.8514170677574	14.0219903546085\\
13.7071130073365	14.1170230578379\\
13.6468196853487	14.1425474628567\\
13.5260278432448	14.1824794759243\\
13.4615916945302	14.1922302652821\\
13.3403524486097	14.2102096527194\\
13.2727717327475	14.2153362455143\\
13.2179607076317	14.2106409342204\\
13.1131680280859	14.1956883110949\\
13.0143975911031	14.1825706170379\\
12.9056965015807	14.1632301139264\\
12.8207538615772	14.1486075906539\\
12.701384064407	14.1361221278949\\
12.642128700728	14.1283163734716\\
12.562377653956	14.1154616368147\\
12.481419914591	14.1138979238973\\
12.3831929058981	14.1176619283094\\
12.3193730571696	14.1277654588295\\
12.1955329955461	14.1519126587761\\
12.1282886938549	14.1644534848161\\
12.0573942344668	14.1760107420806\\
11.9747479897231	14.1882747702517\\
};\label{pgf:Trackloop}

\addplot [color=myred, only marks, mark=otimes, mark options={solid, thick, myred}, forget plot]
  table[row sep=crcr]{%
23.3225784301758	1.56602966785431\\
};

\node[right, align=left, inner sep=0, myred]
at (axis cs:23.323,2.5) {}; 
\addplot [color=myred, only marks, mark=otimes, mark options={solid, thick, myred}, forget plot]
  table[row sep=crcr]{%
8.97374820709229	11.1851282119751\\
};

\node[right, align=left, inner sep=0, myred]
at (axis cs:7.974,10.4) {}; 
\addplot [color=myred, only marks, mark=otimes, mark options={solid, thick, myred}, forget plot]
  table[row sep=crcr]{%
20.9867897033691	11.1696033477783\\
};

\node[right, align=left, inner sep=0, myred]
at (axis cs:19.987,10.4) {}; 
\addplot [color=myred, only marks, mark=otimes, mark options={solid, thick, myred}, forget plot]
  table[row sep=crcr]{%
4.97817087173462	11.1690120697021\\
};

\node[right, align=left, inner sep=0, myred]
at (axis cs:3.978,10.4) {}; 
\addplot [color=myred, only marks, mark=otimes, mark options={solid, thick, myred}, forget plot]
  table[row sep=crcr]{%
7.05806446075439	1.57792556285858\\
};

\node[right, align=left, inner sep=0, myred]
at (axis cs:7.058,2.5) {}; 
\addplot [color=myred, only marks, mark=otimes, mark options={solid, thick, myred}, forget plot]
  table[row sep=crcr]{%
15.0362100601196	1.56918466091156\\
};

\node[right, align=left, inner sep=0, myred]
at (axis cs:15.036,2.5) {}; 
\addplot [color=myred, only marks, mark=otimes, mark options={solid, thick, myred}, forget plot]
  table[row sep=crcr]{%
24.9614124298096	11.1809988021851\\
};

\node[right, align=left, inner sep=0, myred]
at (axis cs:23.961,10.4) {}; 
\addplot [color=myred, only marks, mark=otimes, mark options={solid, thick, myred}, forget plot]
  table[row sep=crcr]{%
27.008768081665	1.56994450092316\\
};

\node[right, align=left, inner sep=0, myred]
at (axis cs:27.009,2.5) {}; 
\addplot [color=myred, only marks, mark=otimes, mark options={solid, thick, myred}, forget plot]
  table[row sep=crcr]{%
19.0649509429932	1.5576354265213\\
};

\node[right, align=left, inner sep=0, myred]
at (axis cs:18.565,2.5) {}; 
\addplot [color=myred, only marks, mark=otimes, mark options={solid, thick, myred}, forget plot]
  table[row sep=crcr]{%
11.0630111694336	1.55558657646179\\
};

\node[right, align=left, inner sep=0, myred]
at (axis cs:11.063,2.5) {}; 
\addplot [color=myred, only marks, mark=otimes, mark options={solid, thick, myred}, forget plot]
  table[row sep=crcr]{%
12.9600801467896	11.1879482269287\\
};

\node[right, align=left, inner sep=0, myred]
at (axis cs:11.96,10.4) {}; 
\addplot [color=myred, only marks, mark=otimes, mark options={solid, thick, myred}]
  table[row sep=crcr]{%
16.9596405029297	11.1709003448486\\
};\label{pgf:antennasIntro}

\node[right, align=left, inner sep=0, myred]
at (axis cs:15.96,10.4) {}; 

\end{axis}
\end{tikzpicture}%
